\documentclass[twocolumn,showpacs,aps,nofootinbib]{revtex4}
\usepackage{amsmath}
\usepackage{epsfig}
\usepackage{subfigure}
\usepackage{float}
\usepackage{verbatim} 

\usepackage{pstricks}
\usepackage{color}
\usepackage{slashed}
\usepackage{multirow}
\usepackage{pstricks}
\usepackage{color}
\usepackage{booktabs}
\usepackage{subfigure}
\usepackage{epstopdf}

\newcommand{\ie}{{\it i.e.~}}

\newcommand{\lag}{{\cal L}}

\renewcommand{\d}[1]{\ensuremath{\operatorname{d}\!{#1}}}

\renewcommand{\lag}[1]{\ensuremath{\mathcal{L}_\mathrm{#1}}}

\newcommand{\FDF}{\left(\varphi^\dagger\overleftrightarrow{D}_\mu\varphi\right)}
\newcommand{\FDFI}{\left(\varphi^\dagger\overleftrightarrow{D}^I_\mu\varphi\right)}

\newcommand{\DET}{\Gamma(1+\epsilon)\left(\frac{4\pi\mu^2}{m_t^2}\right)^\epsilon}
\newcommand{\DE}{\Gamma(1+\epsilon)\left(4\pi\right)^\epsilon}

\renewcommand{\Re}{\ensuremath{\mathrm{Re}}}
\renewcommand{\Im}{\ensuremath{\mathrm{Im}}}
\newcommand{\xhat}{\ensuremath{\hat x}}
\newcommand{\yhat}{\ensuremath{\hat y}}

\graphicspath{{fig/}}

\begin{document}

\leftline{}
\rightline{CP3-14-17}
\title{Effective field theory approach to top-quark decay 
at next-to-leading order in QCD}
\author{Cen Zhang}
\affiliation{
Centre for Cosmology, Particle Physics and Phenomenology, 
Universit\'e Catholique de Louvain, B-1348 Louvain-la-Neuve, Belgium 
}

\begin{abstract} We present analytical results for top-quark decay
  processes, in an effective field theory beyond the Standard Model, at
  next-to-leading order in QCD.  We parametrize new physics effects using
  dimension-six operators, and consider all operators that give rise to
  non-standard interactions of the top quark.  We investigate both the
  flavor-conserving and flavor-changing decay modes, including their two-body
  and three-body semi-leptonic final states.  The QCD mixing among relevant
  operators are also taken into account.  These results provide all information
  needed for a complete model-independent study of top-quark decay at
  next-to-leading order accuracy, paving the way to global analyses for new
  physics effects in an effective field theory approach.
\end{abstract} 


\maketitle


\section{Introduction}
\label{sec:intro}
The discovery of the Higgs boson at the LHC finally completes the Standard
Model (SM) \cite{Aad:2012tfa,Chatrchyan:2012ufa}.  The absence of any resonant
signal of new physics up to several hundreds of GeV allows one to parametrize
the effects of any possible new physics using an effective field theory (EFT)
approach \cite{Weinberg:1978kz,Weinberg:1980wa,Georgi:1994qn}.  In this
approach one assumes that the new physics decouples from the SM in the limit
that the energy scale $\Lambda$, which characterizes the new particles and
interactions goes to infinity, and at the electroweak scale the deviations from
the SM are parameterized by higher-dimensional operators that involve only the
SM fields.  In this paper we also assume that the Higgs boson observed at the
LHC is the SM Higgs boson, and that the electroweak $SU(2)_L\times U(1)_Y$
gauge symmetry is linearly realized.

On the other hand, the top quark continues to be the heaviest particle known,
serving as a window to new physics.  Measurements on top quark production and
decay processes can provide key information on physics beyond the SM. Model
independent analyses of the top quark have been perfermed using the EFT
approach, see for example Refs.~\cite{
AguilarSaavedra:2008gt,AguilarSaavedra:2010rx,AguilarSaavedra:2010sq,AguilarSaavedra:2010zi,AguilarSaavedra:2012vh,Lillie:2007hd,Kumar:2009vs,Grzadkowski:2003tf,Grzadkowski:2005ye,Grzadkowski:1995te,Grzadkowski:1997cj,Cao:2006pu,Cao:2007ea,Whisnant:1997qu,Yang:1997iv,Zhang:2010px,Zhang:2010dr,Greiner:2011tt,Zhang:2012cd,Willenbrock:2012br,Degrande:2010kt,Degrande:2011rt,Degrande:2012gr,Kamenik:2011dk,Shao:2011wa,Kamenik:2011wt,Ko:2012sm,Degrande:2012zj,Atwood:2013xg,Taghavi:2013wy,Hioki:2013rny,Fiolhais:2013fwa,Faller:2013gca,Hayreter:2013kba,Agram:2013koa,Bernreuther:2013aga,Zhang:2013xya,Fabbrichesi:2013bca,Bouzas:2013jha,Adelman:2013jga,Han:2013sea}.

As the precision level of experimental measurements continue to increase, it
becomes more and more important to have theoretical predictions at a same or
better precision level.  As a colored particle, any process involving a
top-quark can potentially suffer from a large uncertainty at the leading order
(LO) accuracy in QCD.  For this reason it is desirable to have predictions in
the EFT framework at a next-to-leading order (NLO) accuracy in QCD.

Fortunately, despite being a ``non-renormalizable theory'', an EFT indeed
provides a framework in which radiative corrections can be consistently
handeled, see for example
Refs.~\cite{Greiner:2011tt,Zhang:2012cd,Mebane:2013cra,Mebane:2013zga,Chen:2013kfa}.
Nevertheless going to the NLO accuracy is not a trivial task, for at least two
reasons.  First, at the NLO accuracy level more operators start to contribute,
and including only a subset of operators is not justified.  Second, operators
involving quark or gluon fields will in general evolve and mix with each other.
As a result extracting information on new physics effects is complicated by the
large number of operators.  In order to determine or constrain their
coefficients, one has to perform a global analysis, taking into account all
available data and all relevant operators, a task that, indeed, needs a
dedicated effort.

The aim of this paper is to give contribution to this goal, by providing the
complete analytical results for top-quark decay processes, in the EFT framework
at NLO in QCD, including all relevant dimension-six operators.  Top-quark
decays provide the best places to probe the weak coupling of top quark
\cite{AguilarSaavedra:2006fy}, as well as to discover its flavor-changing
neutral couplings \cite{AguilarSaavedra:2004wm}.  Previous results on QCD
corrections to top-quark decays are available in the literature.  The $t\to bW$
decay and $W$-helicity fractions in the SM have been computed at NLO
\cite{Fischer:2000kx} and at NNLO \cite{Czarnecki:2010gb}.  Decay of a
polarized top quark has been considered in
Refs.~\cite{Fischer:1998gsa,Fischer:2001gp}.  Ref.~\cite{Drobnak:2010ej} has
studied the same decay mode but with anomalous $tbW$ couplings.  The
flavor-changing decays $t\to u_iV$, where $u_i$ is $u$ or $c$ and $V$ can be
$Z$, $\gamma$ or $g$, have been studied in
Ref.~\cite{Drobnak:2010wh,Drobnak:2010by,Zhang:2010bm}, in terms of
dimension-four and five operators (these operators do not have explicit
$SU(2)_L$ symmetry).  The $t\to u_ih$ decay was originally calculated in
Ref.~\cite{Li:1990cp}, and more recently in \cite{Zhang:2013xya} including the
contributions from flavor-changing color-dipole operators.

In this work we complete the analytical results of top-quark decay in an EFT
framework.  We consider the main decay channel $t\to bW$, the flavor changing
decays $t\to u_iV$ and $t\to u_ih$, as well as their corresponding
semi-leptonic three-body final states.  Apart from checking existing results,
we provide new results, including:
\begin{enumerate}
  \item The contribution of the top color-dipole
    operator $O_{tG}$ in the main decay channel. 
  \item The helicity fraction of $Z$ in
    $t\to u_iZ$.
  \item Differential decay rates for semi-leptonic
    decays (and thus also the finite-width effect of $W$ and $Z$), including
    invariant mass and angular distribution.
  \item Contributions from four-fermion
    operators at NLO. 
\end{enumerate}
We also provide the operator mixing at order $\alpha_s$.  

The paper is organized as follows.  In section \ref{sec:approach} we present
our formalism and explain the calculation strategies.  In section
\ref{sec:mixings} we summarize the $\mathcal{O}(\alpha_s)$ running and mixing
among all relevant operators.  We then give our analytical results in section
\ref{sec:analytical}.  In section \ref{sec:pheno} we discuss some numerical
results, and we summarize in section \ref{sec:conclusion}.

\section{Approach}
\label{sec:approach}
\subsection{Operator list}

Assuming electroweak $SU(2)_L\times U(1)_Y$ gauge symmetry and baryon and
lepton number conservation (see Ref.~\cite{Dong:2011rh} for a discussion about
baryon number violation in top quark decay), the higher-dimensional operators
least suppressed by inverse powers of $\Lambda$ are the ones with mass
dimension six.  They are suppressed by $1/\Lambda^2$.  In this work we only
consider dimension-six operators.  The EFT can be written as
\begin{equation}
  \lag{EFT}=\lag{SM}+\sum_i\frac{C_iO_i+C_i^*O_i^\dagger}{\Lambda^2}+\cdots
\end{equation}
where $O_i$ are the dimension-six operators, and $C_i$ are dimensionless
coefficients.  The ellipsis refers to operators of dimension eight or higher,
which we neglect.  The complete list of dimension-six operators was first given
in Ref.~\cite{Buchmuller:1985jz}, and gradually evolved to an independent
operator basis with 59 operators \cite{Grzadkowski:2010es}.  Here we use the
same basis as in Ref.~\cite{Grzadkowski:2010es}. We will use the following
notation for quark fields:
\begin{flalign}
  &Q:\quad \mbox{3rd-generation left-handed quark doublet}
  \nonumber\\
  &q:\quad \mbox{1st-generation left-handed quark doublet}
  \nonumber\\
  &t:\quad \mbox{right-handed top quark}
  \nonumber\\
  &b:\quad \mbox{right-handed bottom quark}
  \nonumber\\
  &u:\quad \mbox{right-handed up and charm quark}
  \nonumber
\end{flalign}

We first list the operators for two-body decay processes. The relevant
ones for $t\to bW$ are flavor diagonal. They are
\begin{flalign}
  &O_{\varphi Q}^{(3)}
  =i\frac{1}{2}y_t^2 \FDFI (\bar{Q}\gamma^\mu\tau^I Q)
  \label{eq:Ofq3}
  \\
  &O_{tW}=y_tg_W(\bar{Q}\sigma^{\mu\nu}\tau^It)\tilde{\varphi}W_{\mu\nu}^I
  \\
  &O_{\varphi\varphi}=iy_t^2\left(\tilde{\varphi}^\dagger
  D_\mu\varphi\right)(\bar{t}\gamma^\mu b)
  \\
  &O_{bW}=y_tg_W(\bar{Q}\sigma^{\mu\nu}\tau^Ib)\varphi W_{\mu\nu}^I
  \\
  &O_{tG}=y_tg_s(\bar{Q}\sigma^{\mu\nu}T^At)\tilde{\varphi}G_{\mu\nu}^A
  \\
  &O_{t\varphi}=-y_t^3(\varphi^\dagger\varphi)(\bar{Q}t)\varphi
  \label{eq:Otf}
\end{flalign}
where $\tau^I$ are the Pauli matrices, $\varphi$ is the Higgs doublet, and
$\tilde\varphi=i\tau^2\varphi^*$. The covariant derivative $D_\mu$ is defined
as
\begin{flalign}
  D_\mu=\partial_\mu-ig_sT^AG_\mu^A-ig_W\frac{1}{2}\tau^IW_\mu^I-ig_YYB_\mu
  \ .
\end{flalign}
The relevant operators for $t\to u_iV$ and $t\to u_ih$ are flavor off-diagonal.
They are
\begin{flalign}
  &O_{\varphi q}^{(3,1+3)}=i\frac{1}{2}y_t^2\FDFI(\bar{q}\gamma^\mu\tau^IQ)
  \label{eq:Ofq313}
  \\
  &O_{\varphi q}^{(1,1+3)}=i\frac{1}{2}y_t^2\FDF(\bar{q}\gamma^\mu Q) 
  \\
  &O_{\varphi u}^{(1+3)}=i\frac{1}{2}y_t^2\FDF(\bar{u}\gamma^\mu t)
  \label{eq:Ofu13}
  \\
  &O_{uB}^{(13)}=y_tg_Y(\bar{q}\sigma^{\mu\nu}t)\tilde{\varphi}B_{\mu\nu}
  \label{eq:OuB13}
  \\
  &O_{uW}^{(13)}=y_tg_W(\bar{q}\sigma^{\mu\nu}\tau^It)\tilde{\varphi}W^I_{\mu\nu}
  \\
  &O_{uG}^{(13)}=y_tg_s(\bar{q}\sigma^{\mu\nu}T^At)\tilde{\varphi}G^A_{\mu\nu}
  \\
  &O_{u\varphi}^{(13)}=-y_t^3(\varphi^\dagger\varphi)(\bar{q}t)\tilde\varphi
  \label{eq:Ouf13}
\end{flalign}
where superscript $(1+3)$ and $(13)$ denotes the flavor structure.  For
operators with $(13)$ superscript [Eqs.~(\ref{eq:OuB13})-(\ref{eq:Ouf13})], a
similar set of operators with $(31)$ flavor structure can be obtained by
interchanging $(13)\leftrightarrow (31),\ t\leftrightarrow u$ and
$Q\leftrightarrow q$.  The Hermitian conjugation of these $(31)$ operators will
contribute to $t\to u_iV$ or $t\to u_ih$ in a similar way, but with chirality
structures opposite to those from the $(13)$ operators.  On the other hand, for
the operators with superscript $(1+3)$ superscript
[Eqs.~(\ref{eq:Ofq313})-(\ref{eq:Ofu13})], interchanging the first and the
third generation simply gives the Hermitian conjugation of themselves,
therefore needs not to be considered separately.  Replacing the up quark field
with the charm quark field will give the same set of operators for $(2+3)$,
$(23)$ and $(32)$ flavor structures.

 We have normalized these operators by adding factors of $y_t$, $g_W$, $g_Y$
 and $g_s$.  Here $y_t$ is the top-quark Yukawa coupling, $g_W$ is the weak
 gauge coupling, $g_Y$ is the hypercharge gauge coupling, and $g_s$ is the
 strong coupling. More specifically, we attach a $y_t$ for each Higgs field, a $g_W$
 $(g_Y)$ for each $W$ $(B)$ field, and a $g_s$ for each gluon field.  This is
 helpful in determining the order of operator mixing.  In general, the order of
 mixing between two operators depends on the normalization factors of both
 operators, and so its definition itself has some ambiguity.  In this work we
 are interested in NLO effects in QCD, and we consider them as any
 corrections coming from a virtual gluon in a loop or a real gluon emission
 (and gluon splitting in $t\to u_ig$).
 With the above normalization factors, these corrections are
 automatically of order $\mathcal{O}(\alpha_s)$, and thus it is convenient to
 present results using these conventions.  These normalization factors are also
 consistent with the naive dimension analysis \cite{Manohar:1983md}, and
the presented operators are constructed in the following form:
 \begin{flalign}
   f^2\Lambda^2\left( \frac{\psi}{f\sqrt{\Lambda}} \right)^2
   \left( \frac{y_th}{\Lambda} \right)^a
   \left( \frac{D}{\Lambda} \right)^b
   \left( \frac{gX}{\Lambda^2} \right)^c
   \ ,
 \end{flalign}
with $\Lambda\sim 4\pi f$ and $X$ represents any gauge-field strength tensor.
It has been shown in Ref.~\cite{Jenkins:2013zja} that with this convention the
anomalous dimension for the operator coefficients depends on products of powers
of $\lambda/(4\pi)^4$, $g^2/(4\pi)^2$, and $y^2/(4\pi)^2$ as expected.  In this
work we are interested in the $g_s^2/(4\pi)^2$ part, which is related to a
virtual or real gluon correction, and is expected to be important in top-quark
related processes.  The $y_t$ involved in the normalization is defined by the
on-shell top-quark mass:
  \begin{equation}
    y_t=\frac{\sqrt{2}m_t}{v}
    \ ,
  \end{equation}
where $v$ is the vacuum expectation value of the Higgs field.  This is just for
simplicity. As a result, it does not contribute to the anomalous dimension of
the operators at order $\alpha_s$.

Following Refs.~\cite{AguilarSaavedra:2009mx,Grzadkowski:2010es}, we
have defined the Hermitian derivative terms:
    \begin{flalign}
      &\FDF\equiv i\varphi^\dagger\left(D_\mu-\overleftarrow{D}_\mu\right)\varphi
      \\
      &\FDFI\equiv
      i\varphi^\dagger\left(\tau^ID_\mu-\overleftarrow{D}_\mu\tau^I\right)\varphi
      \;.
    \end{flalign}
A relative plus sign between the two terms on the r.h.s.~would give rise to
redundant operators (in the sense that they can be reduced to other operators
like $O_{t\varphi}$ or $O_{u\varphi}^{(13)}$ by equations of motion). The
advantage of using these definitions is that these terms do not involve a Higgs
field, therefore the number of relevant operators in $t\to u_ih$ is reduced.
The flavor diagonal operator $O_{\varphi Q}^{(3)}$ defined in this way is
Hermitian, so we can ignore the imaginary part of its coefficient. Also note
that we always add a Hermitian conjugation of an operator to the Lagrangian,
even if the operator is Hermitian by itself.
  
The operator $O_{t\varphi}$ does not affect the $t\to bW$ decay. We include it
here only because in principle this operator is required as a counterterm to
render the top-leg correction from $O_{tG}$ finite.  In practice this can be
avoided, as we will discuss in Section~\ref{sec:operatorren}.  Similarly, the
operator $O_{u\varphi}^{(13,31)}$ is required as a counterterm to regulate the
one-loop $\bar{u}t$ mixing from $O_{uG}^{(13,31)}$ and from itself.  They are
important in $t\to u_ih$ decay, but can be safely ignored in $t\to u_iV$, as has
been done in previous calculations.

We now list the four-fermion operators, relevant in semi-leptonic decays
\cite{AguilarSaavedra:2010zi}.  These operators have two quark fields and two
lepton fields.  We again choose the operator basis as presented in
Ref.~\cite{Grzadkowski:2010es}, where these operators are Fierzed into the form
of a quark current times a lepton current. This has the advantage that the QCD
correction can be factorized, and for vector and scalar current it is
essentially the same as in the two-body decays, for a given invariant mass of
the lepton pair. This allows us to infer some results from two-body decays.  In
addition, these operators do not mix at order $\alpha_s$, because QCD
correction does not affect the lepton current.

We divide the four-fermion operators into three classes, according to their
Lorentz structures: the vector-vector operator (V-V), the scalar-scalar
operator (S-S), and the tensor-tensor operator (T-T).  The operators that
contribute to flavor-conserving decay $t\to b\nu e^+$ are:

\noindent V-V
\begin{flalign}
  &O_{lQ}^{(3)}=\left( \bar{l}\gamma_\mu\tau^Il \right)
  \left( \bar{Q}\gamma^\mu\tau^IQ \right)
\end{flalign}
S-S
\begin{flalign}
  &O_{lebQ}=\left( \bar{l}e \right)\left( \bar{b}Q \right)
  \\
  &O_{leQt}^{(1)}=\left( \bar{l}e \right)\varepsilon\left(\bar{Q}t\right)
\end{flalign}
T-T
\begin{flalign}
  &O_{leQt}^{(3)}=\left(\bar{l}\sigma_{\mu\nu}e\right)\varepsilon
  \left(\bar{Q}\sigma^{\mu\nu}t\right)
\end{flalign}
where $\varepsilon=i\tau^2$.  The following ones contribute to FCNC decay,
$t\to ul^+l^-$:

\noindent V-V
\begin{flalign}
  &O_{lq}^{(1,1+3)}=\left( \bar{l}\gamma_\mu l \right)
  \left( \bar{q}\gamma^\mu Q \right)
  \\
  &O_{lq}^{(3,1+3)}=\left( \bar{l}\gamma_\mu\tau^I l \right)
  \left( \bar{q}\gamma^\mu\tau^I Q \right)
  \\
  &O_{lu}^{(1+3)}=\left( \bar{l}\gamma_\mu l \right)
  \left( \bar{u}\gamma^\mu t \right)
  \\
  &O_{qe}^{(1+3)}=\left( \bar{q}\gamma_\mu Q \right)
  \left( \bar{e}\gamma^\mu e \right)
  \\
  &O_{eu}^{(1+3)}=\left( \bar{e}\gamma_\mu e \right)
  \left( \bar{u}\gamma^\mu t \right)
\end{flalign}
S-S
\begin{flalign}
  &O_{lequ}^{(1,13)}=\left( \bar{l}e \right)\varepsilon\left( \bar{q}t \right)
  \\
  &O_{lequ}^{(1,31)}=\left( \bar{l}e \right)\varepsilon\left(\bar{Q}u\right)
\end{flalign}
T-T
\begin{flalign}
  &O_{lequ}^{(3,13)}=\left(\bar{l}\sigma_{\mu\nu}e\right)\varepsilon
  \left(\bar{q}\sigma^{\mu\nu}t\right)
  \\
  &O_{lequ}^{(3,31)}=\left(\bar{l}\sigma_{\mu\nu}e\right)\varepsilon
  \left(\bar{Q}\sigma^{\mu\nu}u\right)\ .
\end{flalign}

\subsection{Calculation strategy}

In this work we compute the top decay processes at NLO in QCD but at LO in
$C/\Lambda^2$.  Calculation of higher orders in $C/\Lambda^2$ requires a
complete knowledge of dimension-eight operators, and is beyond the scope of
this paper. For the FCNC decays, LO in $C/\Lambda^2$ means order
$(C/\Lambda^2)^2$ since there is not SM contribution.  For $t\to bW$, with
left-handed or longitudinal polarization of $W$, the LO result is of order
$C/\Lambda^2$. The right-handed helicity vanishes at the tree level in the zero
bottom-quark mass limit, and LO contributions may come from order $(C/\Lambda^2)^2$,
$\alpha_sC/\Lambda^2$, or $Cm_b/\Lambda^2 m_t$.  The non-zero bottom-quark mass
effect at tree level in $t\to bW$ has been given in
Ref.~\cite{AguilarSaavedra:2006fy}. We will take this effect into account when
presenting numerical results, but we will not show the analytical expressions.
 
For $t\to bW$ and $t\to u_iZ$ we compute three helicity states separately for
$W$ and $Z$.  This is done by using the helicity projection operators described
in Ref.~\cite{Fischer:2000kx}.  The helicity fractions can then be used to
derive the differential decay rate of $t\to bW^*\to b\nu e^+$ and $t\to
uZ^*,u\gamma^*\to ul^+l^-$.  By considering three-body final states we are also
including the off-shellness and the finite width effects from $W$ and $Z$.  For
three-body decays mediated by four-fermion S-S and V-V operators, the
differential decay rates can be inferred from two-body results at the NLO
level.  For T-T operators (and their interference with S-S operators), we use a
different set of projection operators to decompose the tensor interaction.  We
discuss this at the end of Section \ref{sec:threebody}.

We use dimensional regularization \cite{'tHooft:1972fi} to regulate both UV and
IR divergences. Whenever $\gamma^5$ is present in our computation, we use the
following prescription based on the 't Hooft-Veltman scheme
\cite{Larin:1993tq,Ball:2004rg}:
\begin{flalign}
  &\gamma^5\to (1-8a_s)\frac{i}{4\!}\varepsilon_{\nu_1\nu_2\nu_3\nu_4}
  \gamma^{\nu_1} \gamma^{\nu_2} \gamma^{\nu_3} \gamma^{\nu_4}
  \\
  &\gamma_\mu\gamma^5\to (1-4a_s)\frac{i}{3!}\epsilon_{\mu\nu_1\nu_2\nu_3}
  \gamma^{\nu_1}\gamma^{\nu_2}\gamma^{\nu_3}
  \\
  &\sigma_{\mu\nu}\gamma^5\to
  -\frac{i}{2}\epsilon_{\mu\nu\alpha\beta}\sigma^{\alpha\beta}
  \ ,
\end{flalign}
where $a_s=C_F\alpha_s/(4\pi)$.

The calculation procedure for each process can be divided into two steps:
\begin{enumerate}
  \item UV-divergent part, which gives rise to operator mixing and RG equations.
  \item UV-finite part, which gives the finite part of the matrix element.
\end{enumerate}

In the first step, we calculate the UV-divergent part arising from the loop
diagrams, and identify the UV counterterms by applying the $\overline{MS}$
scheme and requiring that the UV-divergent terms in the matrix element cancel.
The outcome of this procedure is a set of counterterms for dimension-six
operators. We then proceed to work out the $\mathcal{O}(\alpha_s)$ anomalous
dimension and the RG equations of the operator coefficients. These equations
characterize the $\mathcal{O}(\alpha_s)$ running and mixing of the coefficients
and can be used to evolve them from a higher scale down to the scale of the top
quark mass.  We summarize them in Section~\ref{sec:mixings}.

In the second step we calculate the UV-finite part. The IR-divergences are
canceled by including the real gluon emission (and for $t\to u_ig$ also the
gluon splitting) corrections. The final result is given in terms of the
coefficients of these operators defined at the scale of the top quark mass.
We present these results in Section~\ref{sec:analytical}.

For the flavor-changing operators, we will only consider the $(1+3)$, $(13)$
and $(31)$ structure, keeping in mind that similar results for $t\to cV$ and
$t\to ch$ can be obtained by replacing $C_i^{(1+3),(13),(31)}\to
C_i^{(2+3),(23),(32)}$.

Throughout this paper, we ignore the bottom quark and other light quark masses,
and assume that the CKM matrix is identity.

\subsection{Operator renormalization}
\label{sec:operatorren}
We first discuss a few issues in the
operator renormalization procedure. 
For the SM we use a scheme which subtracts
the massless modes according to $\overline{MS}$,
and the massive ones at zero momentum \cite{Beenakker:2002nc}.
For light quark and gluon we use:
\begin{flalign}
  &\delta Z_2^{(q)}=-\frac{\alpha_s}{3\pi}D_\epsilon\left(\frac{1}{\epsilon_{UV}}
  -\frac{1}{\epsilon_{IR}}\right)
  \\
  &\delta Z_2^{(g)}=-\frac{\alpha_s}{2\pi}D_\epsilon
  \left(\frac{N_f}{3}-\frac{5}{2}\right)
  \left( \frac{1}{\varepsilon_{UV}}-\frac{1}{\varepsilon_{IR}} \right)
  -\frac{\alpha_s}{6\pi}D_\epsilon\frac{1}{\epsilon_{UV}}
  \ ,
\end{flalign}
and for the strong coupling:
\begin{flalign}
  &\delta Z_{g_s}=\frac{\alpha_s}{4\pi}\DE\left(\frac{N_f}{3}-\frac{11}{2}\right)
  \frac{1}{\epsilon_{UV}}+\frac{\alpha_s}{12\pi}D_\epsilon\frac{1}{\epsilon_{UV}}
  \ ,
\end{flalign}
where
\begin{equation}
  D_\epsilon\equiv\DET
  \ .
\end{equation}
We consider five light flavors in the running of $\alpha_s$.  We then apply the
$\overline{MS}$ scheme to the dimension-six operators.  The operators considered
here will only mix with other dimension-six operators, and the counterterms are
given by
\begin{equation}
  C^0_i\to Z_{i,j}C_j=(\mathbf{1}+\delta Z)_{i,j}C_j
  \ .
\end{equation}
In general they could also affect the running of SM parameters
\cite{Jenkins:2013zja}, but this is not an $\mathcal{O}(\alpha_s)$ effect and is
irrelevant for this study.

\begin{figure}[h]
  \begin{center}
    \includegraphics[scale=0.5]{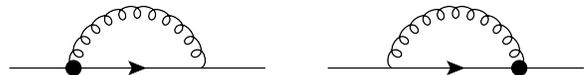}
  \end{center}
  \caption{Top-quark self-energy correction from $O_{tG}.$}
  \label{fig:1}
\end{figure}

Special care needs to be taken for the renormalization of top-quark field.
This is because the operator $O_{tG}$ contribute to the top quark self energy
through the diagrams in Figure~\ref{fig:1}. If one follows the standard
on-shell renormalization condition, the divergence in the mass term will
require a mass counterterm from the operator $ O_{t\varphi}$.  On the other
hand, the wavefunction part needs to be canceled by other operators that are
redundant in our operator basis. This makes the problem more complicated
because these redundant operators will also contribute to $t\to bW$.
Furthermore, since we use $\overline{MS}$ scheme for the operators, only the
pole is canceled in leg corrections, and the remaining finite part  needs to be
taken into account.

The easiest way to avoid the complication, is to replace the
$\varphi^\dagger\varphi$ part in all operators by $\left(
\varphi^\dagger\varphi -v^2/2\right)$. This removes the dimension-four terms in
a dimension-six operator. It corresponds to a redefinition of a SM term, and
thus has no physical effects. The only differences are that some vertices
involving Higgs field will be shifted, and that the two-point counterterms are
provided only by SM terms. Therefore one can adjust the SM counterterms to have
on-shell subtraction including dimension-six terms. The dimension-six
counterterms will then be fixed only by requiring that the matrix element is
finite. 

If we assume no $CP$-violation, then the coefficient $C_{tG}$ is real. Following
the above strategy we obtain the following counterterms for the top quark:
\begin{flalign}
  \delta Z_2^{(t)}=&-\frac{\alpha_s}{3\pi}D_{\epsilon}\left(
  \frac{1}{\varepsilon_{UV}}+\frac{2}{\varepsilon_{IR}}+4\right)
  \nonumber\\
  &+
  \mathrm{Re}C_{tG}\frac{m_t^2}{\Lambda^2}\left(-\frac{2\alpha_s}{\pi}\right)
  D_{\epsilon}
  \left(\frac{1}{\varepsilon_{UV}}+\frac{1}{3}\right)
  \\
  \frac{\delta m_t}{m_t}=&-\frac{\alpha_s}{3\pi}D_{\epsilon}
  \left(\frac{3}{\varepsilon_{UV}} +4\right)
  \nonumber\\
  &+
  \mathrm{Re}C_{tG}\frac{m_t^2}{\Lambda^2}\left(-\frac{4\alpha_s}{\pi}\right)
  D_{\epsilon}
  \left(\frac{1}{\varepsilon_{UV}}+\frac{1}{3}\right)
  \ .
\end{flalign}
These results would apply even if $C_{tG}$ is complex.  In this case there will
be terms proportional to $i\gamma^5$ left in the top-quark self-energy
correction.  They only give a phase to the amplitude of $t\to bW$, which will
not interfere with the SM amplitude, and so these terms have no effect at
$\mathcal{O}(\alpha_s)$.

We use the same strategy to deal with the two-point $u-t$ function arising at
one-loop, from the operator $O_{uG}^{(13,31)}$. We redefine the operator
$O_{u\varphi}^{(13)}$ by
\begin{flalign}
  O_{u\varphi}^{(13)}=-y_t^3\left( \varphi^\dagger\varphi-\frac{v^2}{2} \right)
  \left( \bar{q}t \right)\tilde\varphi
\end{flalign}
and in a similar way for $O_{u\varphi}^{(31)}$. Note the $uth$ vertex is
rescaled by a factor of $2/3$. We then introduce the following $u-t$
counterterms:
\begin{flalign}
  u_{L,0}=\left(1+\frac{1}{2}\delta Z_2^{(q)}\right)u_L
  +\frac{1}{2}\delta Z_{ut}^L t_L
  \\
  u_{R,0}=\left(1+\frac{1}{2}\delta Z_2^{(q)}\right)u_R
  +\frac{1}{2}\delta Z_{ut}^R t_R
  \\
  t_{L,0}=\left(1+\frac{1}{2}\delta Z_2^{(t)}\right)t_L
  +\frac{1}{2}\delta Z_{ut}^L u_L
  \\
  t_{R,0}=\left(1+\frac{1}{2}\delta Z_2^{(t)}\right)t_R
  +\frac{1}{2}\delta Z_{ut}^R u_R
\end{flalign}
with
\begin{flalign}
  &\frac{1}{2}\delta Z_{ut}^L=\frac{\alpha_s}{\pi}\frac{C_{uG}^{(13)}m_t^2}{\Lambda^2}
  D_\epsilon\left( \frac{1}{\varepsilon_{UV}}-\frac{2}{3}-i\pi \right)\\
  &\frac{1}{2}\delta Z_{tu}^{L*}=-\frac{2\alpha_s}{\pi}\frac{C_{uG}^{(13)}m_t^2}{\Lambda^2}
  D_\epsilon\left( \frac{1}{\varepsilon}+\frac{1}{3} \right)\\
  &\frac{1}{2}\delta Z_{ut}^{R}=\frac{\alpha_s}{\pi}\frac{C_{uG}^{(31)*}m_t^2}{\Lambda^2}
  D_\epsilon\left( \frac{1}{\varepsilon}-\frac{2}{3}-i\pi \right)\\
  &\frac{1}{2}\delta Z_{tu}^{R*}=-\frac{2\alpha_s}{\pi}\frac{C_{uG}^{(31)*}m_t^2}{\Lambda^2}
  D_\epsilon\left( \frac{1}{\varepsilon}+\frac{1}{3} \right)
  \ .
\end{flalign}
These counterterms are adjusted such that the two-point $u-t$ function is
exactly canceled both at the up-quark shell and the top-quark shell. This is
convenient because one does not need to include any leg correction diagrams in
flavor-changing processes.

\subsection{Three-body final states}
\label{sec:threebody}

We briefly outline our approach for three-body decays, \ie $t\to b\nu e^+$ and
$t\to ul^+l^-$.  We consider two cases. The first is that a two-fermion
operator contribute to a two-body decay $t\to q+X$, and the latter is followed
by $X\to l^+l^-$. The second case is that a direct contribution comes from a
contact $tqll$ interaction.

The contributions from two-fermion operators can be completely factorized for a
given invariant mass of lepton pair, therefore one can make use of the NLO
two-body decay results to infer the three-body decay rates.  The same approach
is often used in $t$ and $B$ decays, nevertheless we sketch out the steps here,
so that we can easily include any non-standard interactions and four-fermion
operators later on.

Consider $t\to q+W^*$ followed by $W^*\to e^+\nu$ as an example. Here $W^*$ is
an off-shell $W$ with virtuality $Q^2$.  The amplitude of the three-body final
state can be obtained from the two-body ones simply by replacing the $W$
polarization vector with a lepton current
\begin{flalign}
  &M^{(bW)}_\mu(m_W^2)\epsilon^{\mu*}
  \nonumber\\
  &\rightarrow
  M^{(bW)}_\mu(Q^2)\frac{g}{\sqrt{2}}\bar{\nu}(k_1)\gamma^\mu P_L
  e(k_2)D(Q^2,m_W,\Gamma_W)^{-1}
\end{flalign}
where $k_1$ and $k_2$ are the momenta of final state leptons,
$M^{(bW)}_\mu(Q^2)$ is the two-body amplitude with the $W$ mass replaced by the
virtuality of the off-shell $W$, and
\begin{equation}
  D(Q^2,m_W,\Gamma_W)=Q^2-m_W^2+im_W\Gamma_W
  \ .
\end{equation}
Note this procedure can be applied both at LO and NLO in QCD, as the QCD
corrections do not affect lepton current.  $\Gamma_W$ can be a function of
$Q^2$.

The resulting three-body amplitude can further be decomposed with the three
polarization vectors of the virtual $W$:
\begin{flalign}
  M^{(be^+\nu)}=&-\sum_{i=+,0,-}M^{(bW)}_\mu(Q^2) \epsilon_i^{\mu*}\epsilon_i^{\nu}
  \nonumber\\
  &\times
  \frac{g}{\sqrt{2}}\bar{\nu}(k_1)\gamma_\nu P_L
  e(k_2)D(Q^2,m_W,\Gamma_W)^{-1}
  \nonumber\\
  \equiv&
  -\sum_{i=+,0,-}M_i^{(bW)}(Q^2) \frac{g}{\sqrt{2}}L^L_i D(Q^2,m_W,\Gamma_W)^{-1}
\end{flalign}
where $M_i^{(bW)}$ is the two-body amplitude with a specific polarization $i$,
and $L^L_i$ denotes the final state left-handed lepton current polarized to the
same state.  We use ($\theta$,$\phi$) to denote the direction of the $e^+$
three-momentum in the $W$ rest frame, relative to the $W$ momentum in the
top-quark rest frame.  Then one can square the amplitude and integrate over the
$\phi$ angle:
\begin{flalign}
  \int\frac{\d\phi}{4\pi}|M^{be^+\nu}|^2
  &=\left(M_i^{(bW)}(Q^2)M_j^{*(bW)}(Q^2)\right)
  \nonumber\\
  &\times\frac{g^2}{2}|D(Q^2,m_W,\Gamma_W)|^{-2}
  \int\frac{\d\phi}{4\pi}\left( L^L_iL_j^{L*} \right)
  \ .
\end{flalign}
Only the diagonal terms ($i=j$) in $L^L_iL_j^{L*}$ survive after integration:
\begin{flalign}
  \int\frac{\d\phi}{4\pi}\left( L^L_iL_j^{L*} \right)
  =Q^2\delta_{ij}f_i(\cos\theta)
\end{flalign}
where
\begin{flalign}
  &f_+(\cos\theta)=\frac{1}{4}\left( 1+\cos\theta \right)^2
  \\
  &f_0(\cos\theta)=\frac{1}{2}\sin^2\theta
  \\
  &f_-(\cos\theta)=\frac{1}{4}\left( 1-\cos\theta \right)^2
  \ ,
\end{flalign}
so in the end one only needs $|M_i^{(bW)}(Q^2)|^2$, \ie the two-body squared
amplitude for a given $W^*$ helicity.

Next we consider phase space. One can always write a $n-$body phase space as a
$(n-1)-$ body phase space times a two-body phase space. For $n=3$,
\begin{flalign}
  \frac{\d\Phi^{(be^+\nu)}}{\d Q^2}=&\frac{1}{2\pi}\Phi^{(bW)}(Q^2)\times\Phi^{(e^+\nu)}
  \nonumber\\=&\frac{1}{16\pi^2}\Phi^{(bW)}(Q^2)\int\d\cos\theta\int\frac{d\phi}{4\pi}
  \ .
\end{flalign}
The same can be done for $n=4$ \ie $be^+\nu g$ final state, so the real
corrections can be taken into account in the same way. Putting all pieces
together, we have the result for $t\to be^+\nu$ at NLO in terms of $t\to bW$
partial widths:
\begin{flalign}
  &\frac{\d\Gamma_{be^+\nu}}{\d Q^2 \d\cos\theta}
  =
  \frac{Q^2}{16\pi^2}\frac{g^2}{2}|D(Q^2,m_W,\Gamma_W)|^{-2}\times
  \nonumber\\&
  \left[ \Gamma_{bW}^{(+)}f_+(\cos\theta)+
    \Gamma_{bW}^{(0)}f_0(\cos\theta)+
    \Gamma_{bW}^{(-)}f_-(\cos\theta)
    \right]
    \ .
    \label{eq:two2three}
\end{flalign}

In the case of $t\to ul^+l^-$ the situation is more complicated, because of the
interference between $t\to uZ^*$ and $t\to u\gamma^*$ with the same
semi-leptonic final state.  We leave the details to Section
\ref{sec:analytical}, where the analytical results will be given.

The contributions from S-S and V-V types of four-fermion operators can be
incorporated in a same way.  In each case, the amplitude from these
four-fermion operators mimics the amplitude of $t\to qX^* \to ql^+l^-$, at both
LO and NLO, up to an overall factor and an $X$ propagator, where $X$ can be a
scalar or a vector.  Since the propagator is a pure number at any given $Q^2$,
the problem can be converted to the NLO correction to two-body decays, and
previous results can be used, with some operator coefficients shifted to
include the effect of four-fermion operators.\footnote{This implies that the
  corresponding counterterms can be obtained with the same shift, which is
  expected because QCD corrections only affect the quark current.} In this way
the interference between two-body decays and four-fermion operators is
automatically included.

For T-T type of four-fermion operators, and their interference with S-S
operators, the results cannot be inferred from two-body decays, but the
differential decay rates can be computed in a similar way, using a set of
projection operators to decompose the tensor interaction and to convert the
problem to a two-body decay:
\begin{flalign}
  P_{i,j}^{\mu\nu\mu'\nu'}(p,k)=\epsilon_{i}^{\mu\nu}(p,k)\epsilon_{j}^{\mu'\nu'}(p,k)
\end{flalign}
where $p$ is the top-quark momentum, $k$ is the momentum of the lepton-pair
system, and $\epsilon_{i}^{\mu\nu}$ is a complete basis for antisymmetric tensors:
\begin{flalign}
  \sum_i\epsilon_i^{\mu\nu}\epsilon_i^{\mu'\nu'}=\frac{1}{2}\left( 
  g^{\mu\mu'}g^{\nu\nu'}-g^{\mu\nu'}g^{\nu\mu'}
  \right)\ .
\end{flalign}
$\epsilon_{i}^{\mu\nu}$ can be constructed from $k^\mu$ and its three
polarization vectors $\epsilon_i^\mu(k)$.  This effectively allows us to
convert the problem to a two-body decay $t\to q+X^*$, where $X$ is some
fictitious tensor particle which doesn't propagate, similar to what has been
done for $t\to bW^*\to be^+\nu$.  The angular distribution of the lepton-pair
can then be derived from the polarization state of $X$.  The derivation of
$P^{\mu\nu\mu'\nu'}_{i,j}$ is straightforward, but the results are tedious, so
we will not display them here.  

Finally, there is no interference between V-V and S-S or V-V and T-T operators,
due to chirality suppression of the leptons.

\section{Operator mixings}
\label{sec:mixings}
In this section we summarize the operator running and mixing at
$\mathcal{O}(\alpha_s)$ that are relevant in our analysis.  We will also extend
a bit and include all two-quark operators that involve a third-generation
quark.  A complete calculation of the anomalous dimension matrix for all
dimension-six operators can be found in
Ref.~\cite{Jenkins:2013zja,Jenkins:2013wua,Alonso:2013hga}.  Our results are in
agreement with the $\mathcal{O}(\alpha_s)$ terms in the complete results.

The anomalous dimension matrix $\gamma$ is defined such that
\begin{flalign}
  \mu\frac{\d C_i(\mu)}{\d \mu}=\gamma_{ij}C_j(\mu)
  \ .
\end{flalign}

\subsection{Flavor-conserving operators}

We consider bilinear quark operators that involve the third generation quarks
$t$ and $b$.  First of all, the following operators are relevant in our
calculation:
\begin{flalign}
  &O_{tG}=y_tg_s(\bar{Q}\sigma^{\mu\nu}T^At)\tilde{\varphi}G^A_{\mu\nu}
  \\
  &O_{tW}=y_tg_W(\bar{Q}\sigma^{\mu\nu}\tau^It)\tilde{\varphi}W^I_{\mu\nu}
  \\
  &O_{tB}=y_tg_Y(\bar{Q}\sigma^{\mu\nu}t)\tilde{\varphi}B_{\mu\nu}
  \\
  &O_{t\varphi}=-y_t^3(\varphi^\dagger\varphi)(\bar{Q}t)\tilde\varphi
  \ .
  \label{eq:mix_diagonal1}
\end{flalign}
Among these operators $O_{tW}$ and $O_{tG}$ contribute to $t\to bW$ decay.
$O_{t\varphi}$ is needed to absorb the UV divergence in the top mass correction
from $O_{tG}$.  We include $O_{tB}$ only for completeness.  The anomalous
dimension matrix, for
$C_{tG}$, $C_{tW}$, $C_{tB}$ and $C_{t\varphi}$, is
\begin{equation}
  \gamma=\frac{2\alpha_s}{\pi}\left( 
  \begin{array}{ccccc}
    \frac{1}{6}	& 	0	&	0	&	0	\\
   \frac{1}{3}	& \frac{1}{3}	&	0	&	0	\\
   \frac{5}{9}	& 	0	&\frac{1}{3}	&	0	\\
   	-4	& 	0	&	0	&	-1	\\
  \end{array}
  \right)
  \ .
  \label{eq:AD1}
\end{equation}
We can see that the mixing effects are from operator $O_{tG}$, which
renormalizes all the other operators.  Note that the above anomalous dimension
matrix is not ``closed'', in the sense that $O_{tG}$ will be renormalized by
for example $O_{G}=g_sf^{ABC}G_{\mu}^{A\nu}G_{\nu}^{B\rho}G_{\rho}^{C\mu}$ at
order $\alpha_s$ \cite{Cho:1994yu}.

The following four operators are similar to $O_{tG}$, $O_{tW}$, $O_{tB}$ and
$O_{t\varphi}$, and can be obtained by $t\to b$ and $\tilde\varphi\to \varphi$.
\begin{flalign}
  &O_{bG}=y_tg_s(\bar{Q}\sigma^{\mu\nu}T^Ab){\varphi}G^A_{\mu\nu}
  \\
  &O_{bW}=y_tg_W(\bar{Q}\sigma^{\mu\nu}\tau^Ib){\varphi}W^I_{\mu\nu}
  \\
  &O_{bB}=y_tg_Y(\bar{Q}\sigma^{\mu\nu}b){\varphi}B_{\mu\nu}
  \\
  &O_{b\varphi}=-y_t^3(\varphi^\dagger\varphi)(\bar{Q}b)\varphi
  \ .
\end{flalign}
The operator $O_{bW}$ contributes to $t\to bW$ decay, but only at order
$\Lambda^{-4}$ if the bottom-quark mass is ignored.  Their anomalous dimension
matrix is
\begin{equation}
  \gamma=\frac{2\alpha_s}{\pi}\left( 
  \begin{array}{ccccccc}
    \frac{1}{6}	& 	0	&	0	&	0	\\
   \frac{1}{3}	& \frac{1}{3}	&	0	&	0	\\
   -\frac{1}{9}	& 	0	&\frac{1}{3}	&	0	\\
   	0	& 	0	&	0	&	-1	\\
  \end{array}
  \right)
  \ .
  \label{eq:AD2}
\end{equation}
Comparing with Eq.~(\ref{eq:AD1}), there are two different components,
$\gamma_{31}$ and $\gamma_{41}$.  The difference in $\gamma_{31}$ is due to the
different hypercharges of the top quark and the bottom quark, as this term is
proportional to $Y_t+Y_Q$ for $O_{tW}$, but to $Y_b+Y_Q$ for $O_{bW}$, where
$Y_t=2/3$, $Y_b=-1/3$ are the hypercharges of the right handed top quark and
bottom quark, while $Y_Q=1/6$ is the hypercharge of the left-handed quark
doublet.  The difference in $\gamma_{41}$ is simply because we are normalizing
these operators with $y_t$ instead of $y_b$ (which we neglect), and so this
component is suppressed by $(y_b/y_t)^2$.

Finally, the following operators do not have anomalous dimension:
\begin{flalign}
  &O_{\varphi Q}^{(3)}=i\frac{1}{2}y_t^2\FDFI(\bar{Q}\gamma^\mu\tau^IQ)
  \\
  &O_{\varphi Q}^{(1)}=i\frac{1}{2}y_t^2\FDF(\bar{Q}\gamma^\mu Q) 
  \\
  &O_{\varphi t}=i\frac{1}{2}y_t^2\FDF(\bar{t}\gamma^\mu t)
  \\
  &O_{\varphi b}=i\frac{1}{2}y_t^2\FDF(\bar{b}\gamma^\mu b)
  \\
  &O_{\varphi\varphi}=iy_t^2(\tilde\varphi^+D_\mu\varphi)(\bar{t}\gamma^\mu b)
  \ ,
\end{flalign}
due to current conservation.  Here $O_{\varphi Q}^{(3)}$ and
$O_{\varphi\varphi}$ contribute to top-quark decay.

\begin{figure*}[tbh]
  \begin{center}
      \includegraphics[width=\linewidth]{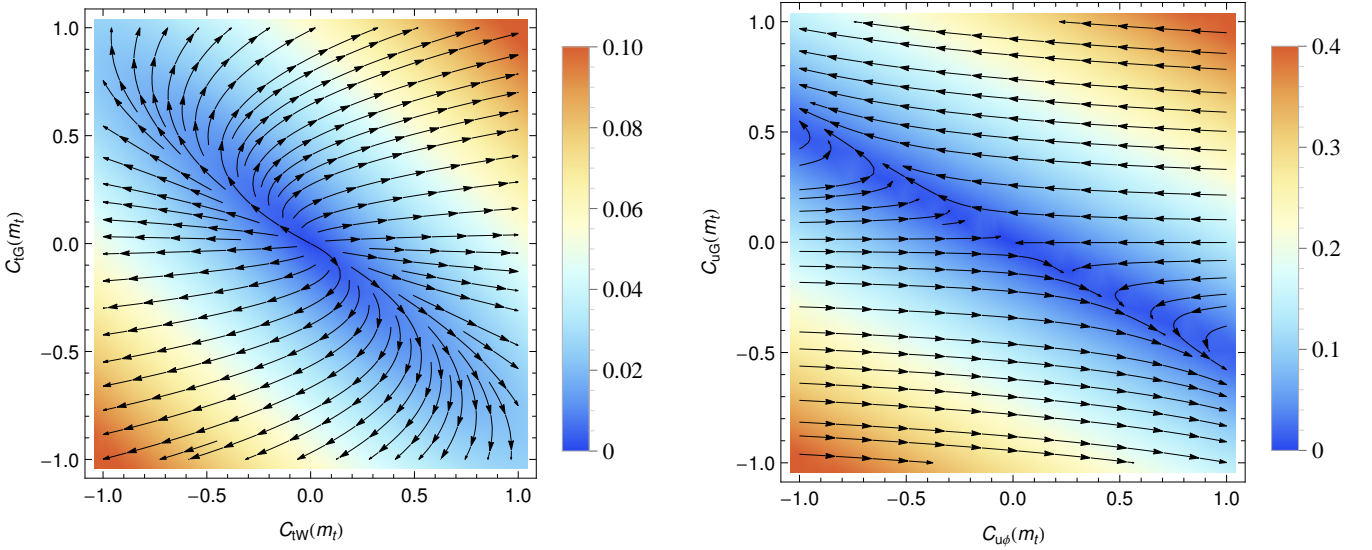}
  \end{center}
  \caption{Left: RG evolution of $C_{tW}$ and $C_{tG}$.  The arrows represent the
  direction of the flow when $\mu$ increases.  The color shows the distance
  between $(C_{tW}(m_t),C_{tG}(m_t))$ and $(C_{tW}(\mu),C_{tG}(\mu))$ for
  $\mu=2$ TeV,
  \ie $\left[\left( C_{tW}(m_t)-C_{tW}(\mu) \right)^2
    +\left( C_{tG}(m_t)-C_{tG}(\mu) \right)^2  \right]^{1/2}$.
  Right: same plot but for $C_{u\varphi}$ and $C_{uG}$.
  \label{fig:rg}
  }
\end{figure*}

To illustrate the mixing effects, we plot the RG evolution of $C_{tW}$ and
$C_{tG}$ in Figure~\ref{fig:rg}.  These two operators affect the $W$-helicity
fractions in $t\to bW$ decay.  Their evolution at order $\mathcal{O}(\alpha_s)$
is determined by the anomalous dimension presented in
Eq.~(\ref{eq:mix_diagonal1}), and is given here by solving the RG equations:
\begin{flalign}
  C_{tW}(\mu)=&
  C_{tW}(m_t)\left( \frac{\alpha_s(\mu)}{\alpha_s(m_t)} \right) ^{-\frac{4}{3\beta_0}}
  \nonumber\\&
  -2C_{tG}(m_t)\left[
  \left( \frac{\alpha_s(\mu)}{\alpha_s(m_t)} \right) ^{-\frac{2}{3\beta_0}}
  -\left( \frac{\alpha_s(\mu)}{\alpha_s(m_t)} \right) ^{-\frac{4}{3\beta_0}}
    \right]
    \\
    C_{tG}(\mu)=&
    C_{tG}(m_t)\left( \frac{\alpha_s(\mu)}{\alpha_s(m_t)} \right) ^{-\frac{2}{3\beta_0}}
    \ ,
  \label{}
\end{flalign}
where $\beta_0=11-2N_f/3$. The second term on the r.h.s of the first equation
implies that there is a mixing effect between these two operators.  The left
plot in Figure~\ref{fig:rg} shows the direction of the RG flows in the
$C_{tW}-C_{tG}$ plane, as well as the distance between
$(C_{tW}(\mu),C_{tG}(\mu))$ at $\mu=m_t$ and at $\mu=2$ TeV.

\subsection{Flavor Changing operators}

We consider bilinear quark operators that involve both first and third
generation quarks.  An interesting feature is that the $\mathcal{O}(\alpha_s)$
mixing will not change the flavor superscripts, due to the chirality of the
massless light quarks.  More specifically, operators with a $(13)$ flavor
indices will not mix with those with $(31)$, and operators with $(1+3)$ are
current operators and do not have any anomalous dimension.  For this reason the
anomalous dimension matrix can be divided into blocks that are closed under RG
evolution at $\mathcal{O}(\alpha_s)$, according to the type and chirality of
the light quark involved in the bilinear quark operator.

\begin{enumerate}
\item{Operators with left-handed massless up quark}
  
The following four operators mix with $O_{uG}^{(13)}$, and
contribute to either $t\to uV$ or $t\to uh$, or both:
\begin{flalign}
  &O_{uG}^{(13)}=y_tg_s(\bar{q}\sigma^{\mu\nu}T^At)\tilde{\varphi}G^A_{\mu\nu}
  \\
  &O_{uW}^{(13)}=y_tg_W(\bar{q}\sigma^{\mu\nu}\tau^It)\tilde{\varphi}W^I_{\mu\nu}
  \\
  &O_{uB}^{(13)}=y_tg_Y(\bar{q}\sigma^{\mu\nu}t)\tilde{\varphi}B_{\mu\nu}
  \\
  &O_{u\varphi}^{(13)}=-y_t^3(\varphi^\dagger\varphi)(\bar{q}t)\tilde\varphi
  \ .
\end{flalign}
Their mixing is
\begin{equation}
  \gamma=\frac{2\alpha_s}{\pi}\left( 
  \begin{array}{ccccccc}
    \frac{1}{6}	&	0	&	0	&	0\\
    \frac{1}{3}	&\frac{1}{3}	&	0	&	0\\
    \frac{5}{9}	&	0	&\frac{1}{3}	&	0\\
    	-2	&	0	&	0	&	-1\\
  \end{array}
  \right)
  \ .
  \label{eq:AD3}
\end{equation}
Comparing with Eq.~({\ref{eq:AD1}}), only the component $\gamma_{41}$ is
different.  This is because $O_{tG}$ is flavor-conserving, and its contribution
to the anomalous dimension comes from two diagrams, each proportional to 
the Yukawa coupling $y_t$, while for $O_{uG}^{(13)}$ one of the two diagrams
would be proportional to $y_u$, which we neglect.

\item{Operators with left-handed massless down quark}
  
The following four operators mix with $O_{dG}^{(13)}$. They are irrelevant for
top decay, but we list them for completeness:
\begin{flalign}
  &O_{dG}^{(13)}=y_tg_s(\bar{q}\sigma^{\mu\nu}T^Ab){\varphi}G^A_{\mu\nu}
  \\
  &O_{dW}^{(13)}=y_tg_W(\bar{q}\sigma^{\mu\nu}\tau^Ib){\varphi}W^I_{\mu\nu}
  \\
  &O_{dB}^{(13)}=y_tg_Y(\bar{q}\sigma^{\mu\nu}b){\varphi}B_{\mu\nu}
  \\
  &O_{d\varphi}^{(13)}=-y_t^3(\varphi^\dagger\varphi)(\bar{q}b)\varphi
  \ .
\end{flalign}
Their mixing is is given by Eq.~(\ref{eq:AD2}).

\item{Operators with right-handed massless up quark}

The following four operators mix with $O_{uG}^{(31)}$, and contribute to either
$t\to uV$ or $t\to uh$, or both:
\begin{flalign}
  &O_{uG}^{(31)}=y_tg_s(\bar{Q}\sigma^{\mu\nu}T^Au)\tilde{\varphi}G^A_{\mu\nu}
  \\
  &O_{uW}^{(31)}=y_tg_W(\bar{Q}\sigma^{\mu\nu}\tau^Iu)\tilde{\varphi}W^I_{\mu\nu}
  \\
  &O_{uB}^{(31)}=y_tg_Y(\bar{Q}\sigma^{\mu\nu}u)\tilde{\varphi}B_{\mu\nu}
  \\
  &O_{u\varphi}^{(31)}=-y_t^3(\varphi^\dagger\varphi)(\bar{Q}u)\tilde\varphi
  \ .
\end{flalign}
Their mixing is is given by Eq.~(\ref{eq:AD3}).

\item{Operators with right-handed massless down quark}

The following four operators mix with $O_{dG}^{(31)}$:
\begin{flalign}
  &O_{dG}^{(31)}=y_tg_s(\bar{Q}\sigma^{\mu\nu}T^Ad){\varphi}G^A_{\mu\nu}
  \\
  &O_{dW}^{(31)}=y_tg_W(\bar{Q}\sigma^{\mu\nu}\tau^Id){\varphi}W^I_{\mu\nu}
  \\
  &O_{dB}^{(31)}=y_tg_Y(\bar{Q}\sigma^{\mu\nu}d){\varphi}B_{\mu\nu}
  \\
  &O_{d\varphi}^{(31)}=-y_t^3(\varphi^\dagger\varphi)(\bar{Q}d)\varphi
  \ .
\end{flalign}
Note the first two contribute to flavor-changing charged-current top decay:
$t\to dW$. Though we didn't study this case explicitly, it is essentially the
same as  $t\to bW$ and the results can be inferred.
Their mixing is given by Eq.~(\ref{eq:AD2}).

\end{enumerate}
Finally, the following operators do not have anomalous dimension:
\begin{flalign}
  &O_{\varphi q}^{(3,1+3)}=i\frac{1}{2}y_t^2\FDFI(\bar{q}\gamma^\mu\tau^IQ)
  \\
  &O_{\varphi q}^{(1,1+3)}=i\frac{1}{2}y_t^2\FDF(\bar{q}\gamma^\mu Q) 
  \\
  &O_{\varphi u}^{(1+3)}=i\frac{1}{2}y_t^2\FDF(\bar{t}\gamma^\mu u)
  \\
  &O_{\varphi d}^{(1+3)}=i\frac{1}{2}y_t^2\FDF(\bar{b}\gamma^\mu d)
  \\
  &O_{\varphi\varphi}^{(13)}=iy_t^2(\tilde\varphi^+D_\mu\varphi)(\bar{t}\gamma^\mu d)
  \\
  &O_{\varphi\varphi}^{(31)}=iy_t^2(\varphi^+D_\mu\tilde\varphi)(\bar{b}\gamma^\mu u)
  \ .
\end{flalign}
The first three contribute to $t\to uZ$.

For illustration we plot the RG evolution of the flavor-changing Yukawa
and color-dipole operators, $O_{u\varphi}$ and $O_{uG}$ (the anomalous dimension
is the same for flavor structure (13) and (31) so we omit the superscript)
in Figure~\ref{fig:rg}.  These two operators will contribute to the decay mode $t\to
u_i h$. Their mixing effect is much larger than that
of $O_{tW}$ and $O_{tG}$.

\subsection{Four-fermion operators with two quarks and two leptons}

Now we consider operators that contribute to semi-leptonic top quark decays.
These operators do not mix at order $\alpha_s$. Their mixing with bilinear quark
operators is not an order $\alpha_s$ effect so we neglect. In the following we
give the anomalous dimensions. The flavor indices do not matter, so we will omit
them. (For example, $O_{lQ}^{(3)}$ and $O_{lq}^{(3,1+3)}$ will have the same
anomalous dimension at $\mathcal{O}(\alpha_s)$, and so here we simply write them
as $O_{lq}^{(3)}$, \textit{etc.})

The V-V operators are not renormalized because of
current conservation. These include
\begin{flalign}
  &O_{lq}^{(1)}=\left( \bar{l}\gamma_\mu l\right)\left( \bar{q}\gamma^\mu q \right)
  \\
  &O_{lq}^{(3)}=\left( \bar{l}\gamma_\mu \gamma^I l\right)
  \left( \bar{q}\gamma^\mu \tau^I q \right)
  \\
  &O_{eu}=\left( \bar{e}\gamma_\mu e\right)\left( \bar{u}\gamma^\mu u \right)
  \\
  &O_{ed}=\left( \bar{e}\gamma_\mu e\right)\left( \bar{d}\gamma^\mu d \right)
  \\
  &O_{lu}=\left( \bar{l}\gamma_\mu l\right)\left( \bar{u}\gamma^\mu u \right)
  \\
  &O_{ld}=\left( \bar{l}\gamma_\mu l\right)\left( \bar{d}\gamma^\mu d \right)
  \\
  &O_{qe}=\left( \bar{q}\gamma_\mu q\right)\left( \bar{e}\gamma^\mu e \right)
  \ .
\end{flalign}

The S-S operators with a scalar quark current have anomalous dimension
\begin{equation}
  \gamma=-2\frac{\alpha_s}{\pi}
  \ .
\end{equation}
These include
\begin{flalign}
  &O_{ledq}=\left( \bar{l}e \right)\left( \bar{d}q\right)
  \\
  &O_{lequ}^{(1)}=\left( \bar{l}e \right)\varepsilon\left( \bar{q}u\right)
  \ .
\end{flalign}

Finally, the T-T operators with a tensor quark current have anomalous dimension
\begin{equation}
  \gamma=\frac{2\alpha_s}{3\pi}
  \ .
\end{equation}
There is only one such operator:
\begin{equation}
  O_{lequ}^{(3)}=\left( \bar{l}\sigma_{\mu\nu}e \right)\varepsilon
  \left( \bar{q}\sigma^{\mu\nu}u \right)\ .
\end{equation}

\section{Analytical results}
\label{sec:analytical}
\subsection{Decay mode: $t\to bW$}
\label{sec:tbW}

In this decay mode, the fractions of $W$ bosons produced with certain
helicities are sensitive to the structure of $tbW$ vertex. Measurements on
helicity fractions can provide information about new physics. In
Ref.~\cite{Drobnak:2010ej}, the QCD corrections to the $W$ boson helicity
fractions with a general anomalous $tbW$ vertex are calculated. Our results on
the partial decay width are in agreement with Ref.~\cite{Drobnak:2010ej}.  Here
we will present these results, and also include one additional operator,
$O_{tG}$. This operator has no tree level contribution and has been ignored in
previous calculations.  However, it gives rise to a chromo-magnetic moment of
the top quark, and thus modifies the standard QCD correction to the SM $t\to
bW$ decay process. It also has a mixing with the other operators, and so needs
to be included.

\begin{figure}[tb]
  \begin{center}
  \includegraphics[scale=0.7]{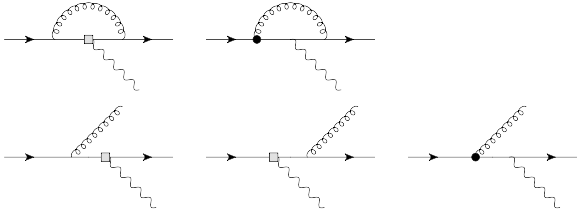}
  \end{center}
  \caption{Virtual and real corrections for $t\to bW$.  The black dots
    represent interactions arising from color-dipole operator $O_{tG}$, while
    squares represent interactions from all the other operators, which modify
    the $tbW$ vertex.  There will be additional diagrams if one includes a
    color-dipole operator for the bottom quark.  These diagrams will not
  interfere with the SM contribution in the limit of $m_b=0$.}
  \label{fig:2}
\end{figure}

In the following we present the full NLO results for top-quark decay to bottom
and $W$ boson in a certain helicity state.  The Feynman diagrams are given in
Figure~\ref{fig:2}.  We give expressions for $\Gamma^{(tot)}$, $\Gamma^{(L)}$,
$\Gamma^{(T)}$ and $\Gamma^{(F)}$, where the superscripts represent the total
width, longitudinal partial width, transverse partial width, and the difference
between transverse positive and transverse negative partial widths.  The decay
width of each helicity state (which we denote by +, - and 0) is thus given by
\begin{flalign}
  &\Gamma^{(+)}=\frac{\Gamma^{(T)}+\Gamma^{(F)}}{2}\\
  &\Gamma^{(0)}=\Gamma^{(L)}\\
  &\Gamma^{(-)}=\frac{\Gamma^{(T)}-\Gamma^{(F)}}{2}
  \ .
\end{flalign}
For later convenience, we write the full results as functions of the $W$-boson
mass and the coefficients of dimension-six operators:
\begin{widetext}
  \begin{flalign}
    \Gamma^{(tot,L,T)}_{bW}\equiv&
    \Gamma^{(tot,L,T)}_{bW}\left(x,C_{\varphi Q}^{(3)},C_{\varphi\varphi},C_{tW},C_{bW}
    ,C_{tG}\right)
    \nonumber\\
    =&\Gamma^{(tot,L,T)}_{\rm SM}(x)
    +\Gamma^{(tot,L,T)}_{1}\left( x,C_{\varphi Q}^{(3)},C_{tW},C_{tG}\right)
    +\Gamma^{(tot,L,T)}_{2}\left( x,C_{\varphi Q}^{(3)},C_{tW}\right)
    +\Gamma^{(tot,L,T)}_{2}\left( x,C_{\varphi\varphi}/2,C_{bW}\right)
    \\
    \Gamma^{(F)}_{bW}\equiv&
    \Gamma^{(F)}_{bW}\left(x,C_{\varphi Q}^{(3)},C_{\varphi\varphi},C_{tW},C_{bW}
    ,C_{tG}\right)
    \nonumber\\
    =&\Gamma^{(F)}_{\rm SM}(x)
    +\Gamma^{(F)}_{1}\left( x,C_{\varphi Q}^{(3)},C_{tW},C_{tG}\right)
    +\Gamma^{(F)}_{2}\left( x,C_{\varphi Q}^{(3)},C_{tW}\right)
    -\Gamma^{(F)}_{2}\left( x,C_{\varphi\varphi}/2,C_{bW}\right)
    \ ,
  \end{flalign}
where $x=m_W/m_t$, $\Gamma_{\rm SM}$ represents the SM contribution, $\Gamma_1$
is the contribution from the interference of SM and dimension-six operators,
and $\Gamma_2$ is the squared contribution from dimension-six operators. We
include the coefficient $C_{tG}$ only in the interference term $\Gamma_1$. A
complete calculation of $\mathcal{O}\left(C_{tG}^2\right)$ effects requires
counterterms from dimension-eight operators (for example, the diagram in
Figure~\ref{fig:1} but with two dots representing $O_{tG}$ may need to be
regulated by a dimension-eight counterterm), and so we will not consider in
this work. The functions
$\Gamma_{\mathrm{SM},1,2}$ are given by
\begin{flalign}
  \label{eq:fulltbw1}
  \Gamma^{(i)}_{\rm SM}\left(x\right)&=
  \frac{\alpha m_t}{4s^2}
  \Gamma_{V,0}^{(i)}\left(x\right)\left( 1+\alpha_s\delta_V^i(x) \right)
  \\
  \Gamma^{(i)}_{1}\left(x,c_V,c_T,c_G\right)&=
  \frac{\alpha m_t^3}{s^2\Lambda^2}\left[
    \Re\left( c_V \right)\Gamma_{V,0}^{(i)}\left(x\right)\left( 1+\alpha_s\delta_V^i(x) \right)
    +\Re\left( c_T \right)\Gamma_{VT,0}^{(i)}\left(x\right)\left( 1+\alpha_s\delta_{VT}^i(x) \right)
    \right.\nonumber\\&\qquad\qquad\qquad\left.
    +\Re\left( c_G \right)\Gamma_{VT,0}\left(x\right)^{(i)}\left(\alpha_s\delta_{VG}^i(x) \right)
    \right]
  \\
  \Gamma^{(i)}_{2}\left(x,c_V,c_T\right)&=
  \frac{\alpha m_t^5}{s^2\Lambda^4}\left[
    \left|c_V\right|^2\Gamma_{V,0}^{(i)}\left(x\right)\left( 1+\alpha_s\delta_V^i(x) \right)
    +\left|c_T\right|^2\Gamma_{T,0}^{(i)}\left(x\right)\left( 1+\alpha_s\delta_T^i(x) \right)
    \right.\nonumber\\&\qquad\qquad\qquad\left.
    +2\Re\left(c_Vc_T^*\right)\Gamma_{VT,0}\left(x\right)^{(i)}\left( 1+\alpha_s\delta_{VT}^i(x) \right)
    \right]
    \ ,
  \label{eq:fulltbw2}
\end{flalign}
\end{widetext}
where $s=\sin\theta_W$ is the sine of the weak angle $\theta_W$, and the
functions $\Gamma^{(i)}_{j,0}(x)$ represent tree level contributions:
\begin{flalign}
  \Gamma_{V,0}^{(tot)}\left(x\right)&=
    \frac{\left(x^2-1\right)^2 \left(2 x^2+1\right)}{4 x^2}
  \label{eq:fulltbw3}
  \\
  \Gamma_{VT,0}^{(tot)}\left(x\right)&=
    \frac{3}{2} \left(x^2-1\right)^2
  \\
  \Gamma_{T,0}^{tot}\left(x\right)&=
    \left(x^2-1\right)^2 \left(x^2+2\right)
\end{flalign}
\begin{flalign}
  \Gamma_{V,0}^{(L)}\left(x\right)&=
\frac{\left(x^2-1\right)^2}{4 x^2}
  \\
  \Gamma_{VT,0}^{(L)}\left(x\right)&=
\frac{1}{2} \left(x^2-1\right)^2
  \\
  \Gamma_{T,0}^{(L)}\left(x\right)&=
x^2 \left(x^2-1\right)^2
\end{flalign}
\begin{flalign}
  \Gamma_{V,0}^{(T)}\left(x\right)&=
  \frac{1}{2} \left(x^2-1\right)^2
  \\
  \Gamma_{VT,0}^{(T)}\left(x\right)&=
  \left(x^2-1\right)^2
  \\
  \Gamma_{T,0}^{(T)}\left(x\right)&=
  2 \left(x^2-1\right)^2
\end{flalign}
\begin{flalign}
  \Gamma_{V,0}^{(F)}\left(x\right)&=
  -\frac{1}{2} \left(x^2-1\right)^2
  \\
  \Gamma_{VT,0}^{(F)}\left(x\right)&=
  -\left(x^2-1\right)^2
  \\
  \Gamma_{T,0}^{(F)}\left(x\right)&=
  -2 \left(x^2-1\right)^2
  \label{eq:fulltbw4}
  \ ,
\end{flalign}
and the functions $\delta^i_{V,T,VT,VG}(x)$ represent $\mathrm{O}(\alpha_s)$
corrections.  Their expressions are given in Appendix~\ref{sec:appNLO}.  Note
that these results apply even if the $W$-boson is off shell, with
$x=m_{W^*}/m_t$.

\subsection{Decay mode $t\to u_iV$}
\label{sec:tuV}

In this section we consider the flavor-changing decay mode $t\to u_iV$ mediated
by dimension-six operators.  In the SM the flavor-changing neutral couplings
involving the top quark are loop-induced and have a strong
Glashow-Iliopoulos-Maiani mechanism suppression, leading to negligible
branching ratios \cite{Eilam:1990zc,Mele:1998ag,AguilarSaavedra:2002ns}.
Therefore the observation of such processes will provide a clear signal of new
physics. 

\begin{figure*}[tb]
  \begin{center}
  \includegraphics[scale=0.75]{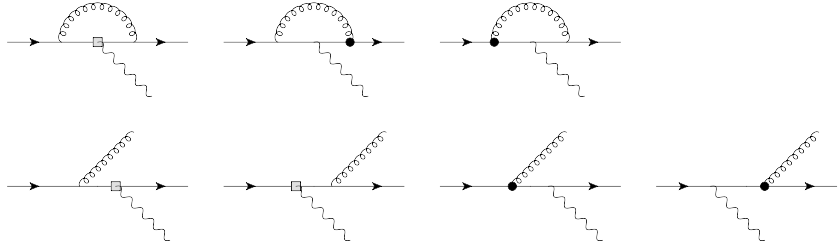}
  \end{center}
  \caption{Feynman diagrams for $t\to u+\gamma/Z$.
    The black dots represent interactions arising from color-dipole operators
    $O_{uG}^{(13,31)}$, while squares represent interactions from all the other
  operators, which modify the $tuV$ vertex.}
  \label{fig:3}
\end{figure*}

\begin{figure*}[tb]
  \begin{center}
  \includegraphics[scale=0.75]{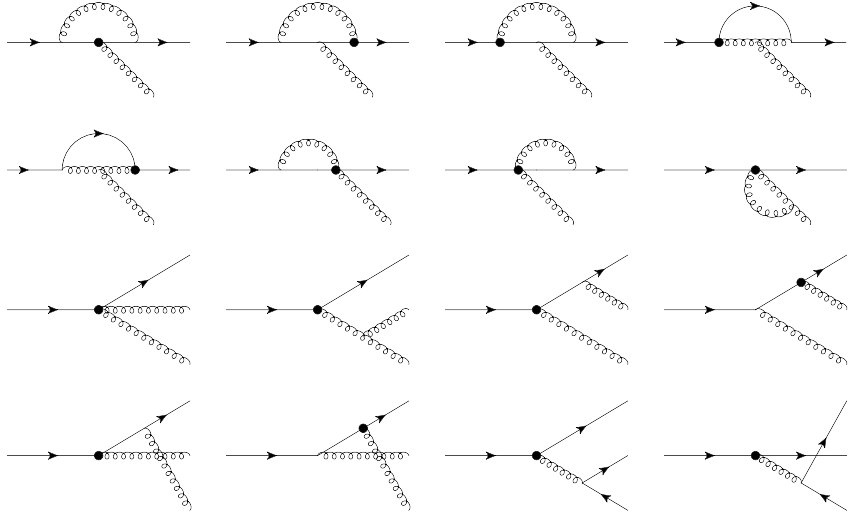}
  \end{center}
  \caption{Feynman diagrams for $t\to u+g$.
    The black dots represent interactions arising from color-dipole operators
    $O_{uG}^{(13,31)}$.}
  \label{fig:4}
\end{figure*}

The QCD corrections have been studied in the literature. In
Ref.~\cite{Zhang:2010bm}, the contributions of dipole couplings to $t \to
u_i+\gamma/Z$ and $t\to u_i+g$ have been investigated at NLO in QCD. The
vector-type couplings for the $tuZ$ vertex are considered in
Refs.~\cite{Drobnak:2010wh,Drobnak:2010by}. Furthermore,
Refs.~\cite{Drobnak:2010wh,Drobnak:2010by} also presented the anomalous
dimensions, in terms of dimension-four and dimension-five effective Lagrangian,
(hence without explicit $SU(2)_L\times U(1)_Y$ symmetry).  For $t \to u_i+Z$,
only the total width, \ie with all polarization states summed over, is
available in these works.  The Feynman diagrams for these processes are shown
in Figure~\ref{fig:3} and \ref{fig:4}.

Here we give the partial decay width of $t\to uZ^*$ for each helicity state of
an off-shell $Z$ boson, in terms of dimension-six operators.  The helicity
decay rates are needed to derive the differential rates of the three-body
decays. Our total widths are in agreement with previous results in
Refs.~\cite{Drobnak:2010wh,Drobnak:2010by}, only if we change the sign of
$\hat{a}$ for the contributions from the $O_{uG}^{(13)}$ operator.  Again, for
later convenience we write the full results as functions of the $Z$-boson mass
and the coefficients of dimension-six operators:
\begin{widetext}
  \begin{flalign}
    \Gamma_{uZ}^{(tot,L,T,F)}\equiv&
    \Gamma_{uZ}^{(tot,L,T,F)}
    \left( x,\hat v,\hat a; C_{\varphi q}^{(1,1+3)}-C_{\varphi q}^{(3,1+3)},
    -s^2C_{uB}^{(13)}+c^2C_{uW}^{(13)},C_{uG}^{(13)};
    C_{\varphi u}^{(1+3)}, -s^2C_{uB}^{(31)*}+c^2C_{uW}^{(31)*},C_{uG}^{(31)*}\right)
    \ ,
  \end{flalign}
  where $\hat v=1/2-4s^2/3$, $\hat a=1/2$ are the vector and axial-vector
  coupling constants of $Zu\bar{u}$, $c=\cos\theta_W$, and the function
  $\Gamma_{uZ}^{(i)}$ is defined as
  \begin{flalign}
    \Gamma_{uZ}^{(tot,L,T)}
    (x,v,a; c_{V}^L,c_{T}^L,c_{G}^L;c_{V}^R,c_{T}^R,c_{G}^R)
    &=\Gamma_{V}^{(tot,L,T)}
    \left(x, v, a; c_V^L, c_T^L,c_G^L\right)
    +\Gamma_{V}^{(tot,L,T)}
    \left(x, v, -a; c_V^R, c_T^R,c_G^R\right)
    \\
    \Gamma_{uZ}^{(F)}
    (x,v,a; c_{V}^L,c_{T}^L,c_{G}^L;c_{V}^R,c_{T}^R,c_{G}^R)
    &=\Gamma_{V}^{(F)}
    \left(x, v, a; c_V^L, c_T^L,c_G^L\right)
    -\Gamma_{V}^{(F)}
    \left(x, v, -a; c_V^R, c_T^R,c_G^R\right)
    \ ,
  \end{flalign}
  where the functions $\Gamma_{V}^{(i)}$ are given by
  \begin{flalign}
    &\Gamma_{V}^{(i)}\left( x,v,a;c_{V},c_{T},c_{G} \right)
    \nonumber\\
    =&\frac{\alpha m_t^5}{2s^2c^2\Lambda^4}\Bigg\{
      \frac{1}{4}\left|c_V\right|^2\Gamma_{V,0}^{(i)}(x)
      \left[ 1+\alpha_s\delta_V^i(x) \right]
      +\left|c_T\right|^2\Gamma_{T,0}^{(i)}(x)
	\left[ 1+\alpha_s\delta_T^i(x) \right]
	-\Re\left( c_V c_T^* \right)\Gamma_{VT,0}^{(i)}
	\left[ 1+\alpha_s\delta_{VT}^i(x) \right]
    \nonumber\\&
    -\Re\left( c_V c_G^* \right)\Gamma_{VT,0}^{(i)}
    \alpha_s\left[ \delta_{VG,r}^i(x,\Re(v),a)-\delta_{VG,i}^i(x,\Im(v),0)\right]
    -\Im\left( c_V c_G^* \right)\Gamma_{VT,0}^{(i)}
    \alpha_s\left[ \delta_{VG,i}^i(x,\Re(v),a)+\delta_{VG,r}^i(x,\Im(v),0)\right]
    \nonumber\\&
    +2\Re\left( c_T c_G^* \right)\Gamma_{T,0}^{(i)}
    \alpha_s\left[ \delta_{TG,r}^i(x,\Re(v),a)-\delta_{TG,i}^i(x,\Im(v),0)\right]
    +2\Im\left( c_T c_G^* \right)\Gamma_{T,0}^{(i)}
    \alpha_s\left[ \delta_{TG,i}^i(x,\Re(v),a)+\delta_{TG,r}^i(x,\Im(v),0)\right]
    \nonumber\\&
      +\left|c_G\right|^2\Gamma_{T,0}^{(i)}(x)
      \alpha_s\left[ \delta_{G2}^i(x,\Re(v),a)+\delta_{G2}^i(x,\Im(v),0)\right]
      \Bigg\}
      \label{eq:fulltuz2}
      \ ,
  \end{flalign}
  where the tree level contributions, $\Gamma_{j,0}^{(i)}$ are given in
  Eq.~(\ref{eq:fulltbw3})-(\ref{eq:fulltbw4}), and the NLO corrections,
  $\delta^{i}(x)$'s, are given in Appendix~\ref{sec:appNLO}.  Note that in the
  above formula we allow for a complex value of $\hat v$, even though $\hat v$
  is real in the SM. This will be useful when we consider four-fermion
  operators in the next section.
\end{widetext}

We have also checked the available results for $t\to u_i g$ in
Ref.~\cite{Zhang:2010bm} and for $t\to u_i\gamma$ in
Ref.~\cite{Drobnak:2010wh,Drobnak:2010by}, and find agreement. For the sake of
completeness, we present these results here using our formalism.

For $t\to u_ig$:
\begin{flalign}
  \Gamma_{ug}=\frac{4\alpha_sm_t^5}{3\Lambda^4}\left( 
  \left|C_{uG}^{(13)}\right|^2+\left|C_{uG}^{(31)}\right|^2\right)
  \left( 1+\alpha_s\delta_G^g \right)
\end{flalign}
with
\begin{flalign}
  \delta_G^g=-\frac{1}{72\pi}\left[ 
    6(29-2N_f)\log\frac{m_t^2}{\mu^2}+36N_f-749+38\pi^2\right]
    \ ,
\end{flalign}
where $N_f=5$ is the number of running flavors in $g_s$.
For $t\to u_i\gamma$:
  \begin{flalign}
    \Gamma_{u\gamma}&=\Gamma_\gamma\left( \hat{x},\hat{y};
    C_{uW}^{(13)}+C_{uB}^{(13)}, C_{uG}^{(13)}\right)
    \nonumber\\&
    +\Gamma_\gamma\left( \hat{x},\hat{y};
    C_{uW}^{(31)*}+C_{uB}^{(31)*}, C_{uG}^{(31)*}\right)
    \ ,
  \end{flalign}
  where $\hat{x}$ and $\hat{y}$ are kinematic cuts on the photon energy and
  the jet-photon separation, required to remove the photon soft-collinear
  divergences:
  \begin{flalign}
    1-\mathbf{p}_\gamma\cdot\mathbf{p}_{u_i}/E_\gamma E_{u_i}>\hat{x}
    \\
    2E_\gamma/m_t>\hat{y}
    \ ,
  \end{flalign}
  where the energy $E_\gamma$, $E_{u_i}$ and three-momenta $\mathbf{p}_\gamma$,
  $\mathbf{p}_{u_i}$ are defined in the top-quark rest frame, 
  see \cite{Drobnak:2010by} for details.  The function $\Gamma_\gamma$ is defined as
\begin{widetext}
  \begin{flalign}
    &\Gamma_\gamma(\xhat,\yhat;c_T,c_G)
    \nonumber\\
    =&\frac{\alpha m_t^5}{\Lambda ^4}
    \Bigg\{|c_T|^2\left[ 1+\alpha_s\delta_T^\gamma(\xhat,\yhat) \right]
      +2\Re\left(c_Tc_G^*\right)\alpha_s\delta_{TG,r}^\gamma(\xhat,\yhat)
      +2\Im\left(c_Tc_G^*\right)\alpha_s\delta_{TG,i}^\gamma(\xhat,\yhat)
      +|c_G|^2\alpha_s\delta_{G}^i(\xhat,\yhat)
    \Bigg\}
    \ ,
    \label{eq:fulltua2}
  \end{flalign}
\end{widetext}
where the NLO corrections, $\delta^\gamma_i(\xhat,\yhat)$, are given in
Appendix \ref{sec:appNLO}.  Note that in both $t\to uZ$ and $t\to u\gamma$, the
contributions from color-dipole operators $O_{uG}^{(13,31)}$ are pure NLO
effects.

\subsection{Decay mode: $t\to u_ih$}\label{sec:tuh}

The QCD correction to $t\to u_ih$ decay through Yukawa operators was first
computed in Ref.~\cite{Li:1990cp}.  The process considered there is the charged
Higgs decay of the top quark, but the QCD correction is the same (after taking
into account the difference between renormalization schemes).  More recently we
presented a calculation for $t\to u_ih$ in Ref.~\cite{Zhang:2013xya}, adding
the contributions from the color-dipole operators $O_{uG}^{(13,31)}$, and their
interferences with the Yukawa operators $O_{u\varphi}^{(13,31)}$. The Feynman
diagrams are listed in Figure~\ref{fig:5}.  For completeness we present here
the full results for this decay mode:
\begin{figure*}[tb!]
  \begin{center}
  \includegraphics[scale=0.8]{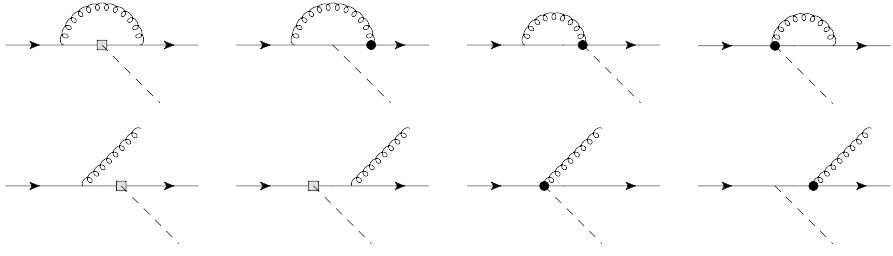}
  \end{center}
  \caption{Virtual and real corrections for $t\to uh$.
    The black dots represent interactions arising from color-dipole operators
    $O_{uG}^{(13,31)}$, while squares represent interactions from Yukawa
    operators $O_{u\varphi}^{(13,31)}$.}
  \label{fig:5}
\end{figure*}
\begin{flalign}
  \Gamma_{uh}=&\Gamma_S\left( x,C_{u\varphi}^{(13)},C_{uG}^{(13)} \right)
  \nonumber\\
  &+\Gamma_S\left( x,C_{u\varphi}^{(31)},C_{uG}^{(31)} \right)
  \ ,
\end{flalign}
where the function $\Gamma_S$ is defined as
\begin{widetext}
\begin{flalign}
  \Gamma_S(x,c_S,c_G)\equiv \left( \frac{G_Fm_t^7}{4\sqrt{2}\pi\Lambda^4} \right)
  \Bigg\{
    \left|c_S\right|^2 \Gamma_{S,0}(x)\left[1+\alpha_s\delta_S(x)\right]
    +2\Re\left( c_Sc_G^* \right) \Gamma_{S,0}(x)\left[\alpha_s\delta_{SG}(x)\right]
    +\left|c_G\right|^2 \Gamma_{S,0}(x)\left[\alpha_s\delta_{G3}(x)\right]
  \Bigg\}
  \ ,
  \label{eq:fulltuh2}
\end{flalign}
\end{widetext}
where $x=m_h/m_t$, and
\begin{equation}
  \Gamma_{S,0}(x)\equiv\left( 1-x^2 \right)^2
\end{equation}
represents the LO contribution.  The $\delta_{S,SG,G3}(x)$ functions are the NLO
corrections, and are given in Appendix~\ref{sec:appNLO}.  Again, here the
contributions the color-dipole operators are pure NLO effects.

\subsection{Three-body final state}

In this section we present results for three-body final states, \ie $t\to
be^+\nu$ and $t\to ul^+l^-$.  The final state leptons will have the same
chirality if the contributions come from two-fermion operators or V-V operators,
or opposite chiralities if they come from S-S or T-T operators.  Since there is
no interference between the two cases, the most convenient way of presenting our
results is to consider them separately.

We first present results for same chirality leptons.  We will give expressions
for the differential decay rate, $\d\Gamma/\d Q^2\d\cos\theta$, in terms of the
two-body NLO results.  Here $Q^2$ is the invariant mass of the lepton pair, and
$\theta$ is the angle between the three-momentum of the anti-lepton in the $W$
rest frame and the $W$ momentum in top-quark rest frame.  We will make use of
the functions $\Gamma_{bW}$ and $\Gamma_{uV}$ defined in section \ref{sec:tbW}
and \ref{sec:tuV}.

For the charged-current decay $t\to be^+\nu$, the result follows immediately
from section~\ref{sec:threebody}:
\begin{flalign}
  &\frac{\d\Gamma_{be^+\nu}}{\d Q^2 \d\cos\theta}
  =
  \frac{Q^2}{16\pi^2}\frac{g^2}{2}|D(Q^2,m_W,\Gamma_W)|^{-2}\times
  \nonumber\\&
  \sum_{i=+,0,-} \Gamma^{(i)}_{bW}
  \left( x, {C_{\varphi Q}^{(3)}}', C_{\varphi\varphi}, C_{tW}, C_{bW}, C_{tG} \right)
  f_i(\cos\theta)
  \ ,
\end{flalign}
where
\begin{equation}
  x^2=Q^2/m_t^2
\end{equation}
and
\begin{flalign}
  {C_{\varphi Q}^{(3)}}'= C_{\varphi Q}^{(3)}
  + \frac{4}{m_t^2g^2} C_{lQ}^{(3)} D(Q^2,m_W,\Gamma_W)
  \ .
\end{flalign}
The above substitution takes into account the four-fermion operator
$O_{lq}^{(3)}$.

Now we turn to the neutral current case. This can be derived from the two-body
decay results for $t\to uZ^*$ and $t\to u\gamma^*$, where $t\to uZ^*$ has been
given in the previous section, and $t\to u\gamma^*$ can be written as:
\begin{widetext}
  \begin{equation}
    \Gamma^{(+,0,-)}_{u\gamma^*}
    =\Gamma^{(+,0,-)}_{uZ}\left(Q^2,\frac{4}{3}sc,0;
    0,sc(C_{uB}^{(13)}+C_{uW}^{(31)}),C_{uG}^{(13)};
    0,sc(C_{uB}^{(31)*}+C_{uW}^{(31)*}),C_{uG}^{(31)*}
    \right)
    \ .
    \label{eq:gamma2Z}
  \end{equation}
\end{widetext}

However the situation is complicated by the interference between $t\to
u\gamma^*$ and $t\to uZ^*$, due to their common semi-leptonic final state.  The
factorization of the decay amplitude can be done in a similar way as for $t\to
bW^*$, but one needs to consider the sum of $M^{(uZ)}$ and $M^{(u\gamma)}$:
\begin{widetext}
  \begin{eqnarray}
    M^{(ul^+l^-)}&=&
    -\sum_{i=+,0,-}\left[
      M_i^{(uZ)}(Q^2) \frac{g}{c}\left( T_3^l-s_W^2Q^l \right)D^{-1}(Q^2,m_Z,\Gamma_Z)
      +M_i^{(u\gamma)}(Q^2) eQ^l D^{-1}(Q^2,0,0)\right]
      \times L^L_i 
    \nonumber\\
    &&-\sum_{i=+,0,-}\left[
      M_i^{(uZ)}(Q^2) \frac{g}{c}\left(-s_W^2Q^l \right)D^{-1}(Q^2,m_Z,\Gamma_Z)
      +M_i^{(u\gamma)}(Q^2) eQ^l D^{-1}(Q^2,0,0)\right]
      \times L^R_i 
  \ ,
  \end{eqnarray}
\end{widetext}
where $T_3^l$ and $Q^l$ are the isospin and electric charge of the final state
leptons $l^+l^-$, and $L_i^{L,R}$ are the polarized lepton currents. Note that
there is no interference between $L^L_i$ and $L^R_i$, and the square of $L^R_i$
is
\begin{flalign}
  \int\frac{\d\phi}{4\pi}\left( L^R_iL_j^{R*} \right)
  =Q^2\delta_{ij}f_i(-\cos\theta)
  \ .
\end{flalign}

For the final state with left-handed leptons, one needs the square of the
combination
\begin{flalign}
    M_i^{(uZ)}(Q^2) 
    +M_i^{(u\gamma)}(Q^2) \frac{eQ^l D(Q^2,m_Z,\Gamma_Z)}
    {\frac{g}{c}\left( T_3^l-s_W^2Q^l \right)D(Q^2,0,0)}
    \ .
\end{flalign}
For the right-handed leptons the same is true if we set $T_3^l=0$.  There is no
need to compute the interference between $M_i^{(uZ)}$ and $M_i^{(u\gamma)}$: we
can write $M_i^{(u\gamma)}$ in terms of $M_i^{(uZ)}$, as in
Eq.~(\ref{eq:gamma2Z}), and directly combine the couplings. This is because
$M_i^{(uZ)}$ is a linear function of $vC_{uG}^{(13,31)}$, $aC_{uG}^{(13,31)}$ ,
and other operators coefficients. In the meantime, the contributions from the
four-fermion operators can be included by a shift in suitable couplings.  The
final result is
\begin{widetext}
\begin{flalign}
  \frac{\d\Gamma_{ul^+l^-}}{\d Q^2\d\cos\theta}
  =&\frac{Q^2}{16\pi^2}|D(Q^2,m_Z,\gamma_Z)|^{-2}\times
  \nonumber\\&
  \Bigg\{
    \left[ \frac{g}{c}\left( T_3^l-s_W^2Q^l \right) \right]^2
    \sum_{i=+,0,-}f^i(\cos\theta)\times
    \Gamma_{uZ}^{(i)}
    (x,v',a; c_{V}^{L\prime},c_{T}^{L\prime},c_{G}^L;
    c_{V}^{R\prime},c_{T}^{R\prime},c_{G}^R)
      \nonumber\\& +
    \left[\frac{g}{c}\left(-s_W^2Q^l \right) \right]^2
    \sum_{i=+,0,-}f^i(-\cos\theta)\times
    \Gamma_{uZ}^{(i)}
    (x,v'',a; c_{V}^{L\prime\prime},c_{T}^{L\prime\prime},c_{G}^L;
    c_{V}^{R\prime\prime},c_{T}^{R\prime\prime},c_{G}^R)
  \Bigg\}
   \ ,
\end{flalign}
\end{widetext}
where $x^2=Q^2/m_t^2$, and
\begin{flalign}
  c_{V}^{L\prime}&=
  C_{\varphi q}^{(1,1+3)}-C_{\varphi q}^{(3,1+3)}
  -\frac{2c^2}{m_t^2g^2}\frac{D(Q^2,m_Z,\Gamma_Z)}{T_3^l-s_W^2Q^l}
  \nonumber\\&\qquad\times
  \left(C_{lq}^{(1,1+3)}+2T_3^lC_{lq}^{(3,1+3)}\right)
  \\
  c_{V}^{R\prime}&=
  C_{\varphi u}^{(1+3)}
  -\frac{2c^2}{m_t^2g^2}\frac{D(Q^2,m_Z,\Gamma_Z)}{T_3^l-s_W^2Q^l}
  C_{lu}^{(1+3)}
  \\
  c_T^{L\prime}&=-s^2C_{uB}^{(13)}+c^2C_{uW}^{(13)}
  +r_L(Q^2)\left(  C_{uB}^{(13)}+   C_{uW}^{(13)}\right)
  \\
  c_T^{R\prime}&=-s^2C_{uB}^{(31)*}+c^2C_{uW}^{(31)*}
  +r_L(Q^2)\left(  C_{uB}^{(31)*}+   C_{uW}^{(31)*}\right)
\end{flalign}
\begin{flalign}
  c_{V}^{L\prime\prime}&=
  C_{\varphi q}^{(1,1+3)}-C_{\varphi q}^{(3,1+3)}
  -\frac{2c^2}{m_t^2g^2}\frac{D(Q^2,m_Z,\Gamma_Z)}{-s_W^2}Q^l
  C_{qe}^{(1+3)}
  \\
  c_{V}^{R\prime\prime}&=
  C_{\varphi u}^{(1+3)}
  -\frac{2c^2}{m_t^2g^2}\frac{D(Q^2,m_Z,\Gamma_Z)}{-s_W^2}Q^l
  C_{eu}^{(1+3)}
  \\
  c_T^{L\prime\prime}&=-s^2C_{uB}^{(13)}+c^2C_{uW}^{(13)}
  +r_R(Q^2)\left(  C_{uB}^{(13)}+   C_{uW}^{(13)}\right)
  \\
  c_T^{R\prime\prime}&=-s^2C_{uB}^{(31)*}+c^2C_{uW}^{(31)*}
  +r_R(Q^2)\left(  C_{uB}^{(31)*}+   C_{uW}^{(31)*}\right)
\end{flalign}
\begin{flalign}
  c_G^L&=C_{uG}^{(13)}
  \\
  c_G^R&=C_{uG}^{(31)*}
  \\
  {v}'&=\frac{1}{2}-\frac{4}{3}s_W^2+\frac{4}{3}r_L(Q^2)
  \\
  {v}''&=\frac{1}{2}-\frac{4}{3}s_W^2+\frac{4}{3}r_R(Q^2)
  \\
  {a}&=\frac{1}{2}
  \\
  r_L(Q^2)&=\frac{s^2c^2Q^l}{T_3^l-s^2Q^l}\frac{D(Q^2,m_Z,\Gamma_Z)}{D(Q^2,0,0)}
  \\
  r_R(Q^2)&=-c^2\frac{D(Q^2,m_Z,\Gamma_Z)}{D(Q^2,0,0)}
  \ .
\end{flalign}
This formula includes contributions from two-fermion operators with $ut\gamma$,
$utZ$ and $utg$ couplings, and from four-fermions operators, as well as their
interferences.

Now we move on to the case where the final state leptons have opposite
chiralities.

The scalar mediated case is straightforward, as
there is no angular correlation
between $t\to uh^*$ and $h^*\to X$, where $X$ is some Higgs decay final state. In
general, if $X$ does not involve any colored state, one can factorize the decay
rate:
\begin{flalign}
  \frac{\d\Gamma_{uX}}{\d Q^2}=&
  \frac{\sqrt{Q^2}}{\pi}\Gamma_{t\to uh^*}\Gamma_{h^*\to X}
  |D(Q^2,m_h,\Gamma_h|^{-2}
  \ .
\end{flalign}
There is no angular distribution in the $h^*$ rest frame.  Furthermore, for
semi-leptonic final states, there is no contribution from two-body decay,
because the Higgs does not couple to leptons in the massless limit. Thus we only
need to consider the contributions from four-fermion operators, S-S operators
$O_{lebQ}$, $O_{leQt}^{(1)}$, $O_{lequ}^{(1,13)}$ and $O_{lequ}^{(1,31)}$, and
T-T operators $O_{leQt}^{(3)}$, $C_{lequ}^{(3,13)}$ and $C_{lequ}^{(3,31)}$.
Their anomalous dimensions are $-2\alpha_s/\pi$ and $2\alpha_s/(3\pi)$
respectively, and there is no mixing among them.  There are also contributions
from the interference of S-S and T-T operators, \ie the interferences between
operators $O_{leQt}^{(3)}$ and $O_{leQt}^{(1)}$, $O_{lequ}^{(3,13)}$ and
$O_{lequ}^{(1,13)}$, and $O_{lequ}^{(3,31)}$ and $O_{lequ}^{(1,31)}$.  Other
interferences vanish due to zero $b$ or $u$ quark mass.

The full results from these four-fermion operators are: 
\begin{widetext}
$t\to be^+\nu$:
\begin{flalign}
  \frac{\d\Gamma_{be^+\nu}}{\d Q^2\d\cos\theta}&=
  \Gamma_{S+T}\left(x,
  \cos\theta; C_{leQt}^{(1)},C_{leQt}^{(3)} \right)
  +\Gamma_{S+T}\left(x,
  \cos\theta; C_{lebQ},0 \right)
  \ ,
\end{flalign}
$t\to ue^+e^-$:
\begin{flalign}
  \frac{\d\Gamma_{ue^+e^-}}{\d Q^2\d\cos\theta}&=
  \Gamma_{S+T}\left(x,
  \cos\theta; C_{lequ}^{(1,13)},C_{lequ}^{(3,13)} \right)
  +\Gamma_{S+T}\left(x,
  \cos\theta;C_{lequ}^{(1,31)*},C_{lequ}^{(3,31)*} \right)
  \ ,
\end{flalign}
where $x^2=Q^2/m_t^2$, and there is no contributions to $t\to u\nu\bar\nu$
because a scalar or a tensor current always involves right-handed neutrino.
The function $\Gamma_{S+T}$ is defined as:
\begin{flalign}
  \Gamma_{S+T}\left( x,\cos\theta; c^{(1)},c^{(3)} \right)
  =&\frac{1}{128\pi^3}\frac{m_t^3}{\Lambda^4}\Bigg\{
    \left|c^{(1)}\right|^2 \gamma_{SS,0}(x,\cos\theta)
    \left[ 1+\alpha_s\delta_S(x) \right]
    \nonumber\\&\quad
    +\Re\left(c^{(1)}c^{(3)*}\right) \gamma_{ST,0}(x,\cos\theta)
    \left[ 1+\alpha_s\delta_{ST}(x) \right]
    +\left|c^{(3)}\right|^2 \sum_{i=1,2}\gamma_{TT,0}^{(i)}(x,\cos\theta)
    \left[ 1+\alpha_s\delta_{TT}^{(i)}(x) \right]
  \Bigg\}
  \ ,
  \label{eq:fullST2}
\end{flalign}
where the tree level contributions are given by
\begin{flalign}
  \gamma_{SS,0}(x,\cos\theta)=&
  \frac{1}{8}x^2\left( 1-x^2 \right)^2
  \left[ f^+(\cos\theta) + f^0(\cos\theta) + f^-(\cos\theta)\right]
  \\
  \gamma_{ST,0}(x,\cos\theta)=&
  -x^2\left( 1-x^2 \right)^2
  \left[ f^+(\cos\theta) - f^-(\cos\theta)\right]
  \\
  \gamma_{TT,0}^{(1)}(x,\cos\theta)=&
  2x^2\left( 1-x^2 \right)^2
  \left[ f^+(\cos\theta) - f^0(\cos\theta) + f^-(\cos\theta)\right]
  \\
  \gamma_{TT,0}^{(2)}(x,\cos\theta)=&
  4\left( 1-x^2 \right)^2
  \left[ f^0(\cos\theta)\right]
  \ ,
\end{flalign}
and the NLO corrections, $\delta_{S,ST}(x)$ and $\delta_{TT}^{(i)}(x)$ are given in
Appendix~\ref{sec:appNLO}.
\end{widetext}

\section{Numerical results}
\label{sec:pheno}
In this section we present some numerical results.  In particular we focus on
the effects of the top-quark color-dipole operator $O_{tG}$ and the
four-fermion operators, which are often ignored in previous studies.
Nevertheless, we will not go into too much detail.  The main purpose of this
paper is to provide analytical results, rather than to discuss their
phenomenological aspects.  A more complete study of their phenomenological
implications, including strategies for searching and bounding the operators,
will presented elsewhere.

Throughout this section we use the following values as input parameters:
\begin{flalign}
  &m_W=80.385\ \mathrm{GeV}\\
  &m_Z=91.1876\ \mathrm{GeV}\\
  &m_t=173 \ \mathrm{GeV}\\
  &m_h=125 \ \mathrm{GeV}\\
  &G_F=1.1663787\times10^{-5}\ \mathrm{GeV}^{-2}\\
  &\alpha_s(m_t)=0.1081
  \ .
\end{flalign}
The strong coupling $\alpha_s(m_t)$ is obtained with RunDec \cite{Chetyrkin:2000yt}
from the value $\alpha_s(m_Z)=0.1185$ \cite{Beringer:1900zz}.

\subsection{$t\to bW$}

The main channel decay has been computed in Ref.~\cite{Drobnak:2010ej} at NLO
and the width and $W$-helicity fractions are given in terms of the anomalous
$tbW$ couplings.  These anomalous couplings are in one-to-one correspondence
with the coefficients of the four operators, $O_{\varphi Q}^{(3)}$,
$O_{\varphi\varphi}$, $O_{tW}$ and $O_{bW}$.  Our results include the
contribution from the top-quark dipole operator $O_{tG}$, which is a pure NLO
effect.  To illustrate its numerical impact, here we will focus only on
$O_{tW}$ and $O_{tG}$.  The other operators are either tightly constrained from
$B$ meson decay \cite{Grzadkowski:2008mf} or do not change the $W$-helicity
fractions.  For the numerical results we will also include the $m_b\neq0$
effects for the dimension-six operators \cite{AguilarSaavedra:2006fy} at LO.
We also include the NNLO QCD correction to the SM contribution
\cite{Czarnecki:2010gb}.  The off-shellness and finite-width effect of the $W$
is taken into consider.  We assume $\Lambda=1$ TeV and both $C_{tW}$ and
$C_{tG}$ are real. 

Up to order $\alpha_s/\Lambda^2$, we find the total width:
\begin{flalign}
  \Gamma^{tot}=&\left[ 1.311+0.158C_{tW}+\alpha_s\left( -0.11C_{tW}-0.04C_{tG} \right)
    \right]\,\mathrm{GeV}
    \nonumber\\
    =&\left(  1.311 +0.146C_{tW}-0.004C_{tG}\right)\,\mathrm{GeV}
    \ ,
\end{flalign}
and the helicity fractions:
\begin{flalign}
  F_0=&0.689-0.040C_{tW}+\alpha_s\left( 0.006C_{tW}+0.007C_{tG} \right)
  \nonumber\\
  =&0.689-0.039C_{tW}+0.0007C_{tG}
  \\
  F_+=&\left[ 1.69-0.04C_{tW}+\alpha_s\left( -0.57C_{tW}+0.31C_{tG} \right)
    \right]\times10^{-3}
    \nonumber\\
    =&\left[ 1.69-0.10C_{tW}+0.03C_{tG} \right]\times10^{-3}
    \ ,
    \label{eq:Fplus}
\end{flalign}
where $F_{0,+}\equiv \Gamma^{(0,+)}/\Gamma^{(tot)}$ are the fractions of
longitudinal and transverse-positive $W$.  The contributions from the non-zero
bottom-quark mass are about $-0.4\%$ for $\Gamma^{tot}$, $-0.1\%$ for $F_{0}$
and $25.7\%$ for $F_{+}$.  We have fixed the renormalization scale at
$\mu=m_t$.

For $\Gamma$ and $F_0$ the influence from $C_{tG}$ is very small, as expected.
It has the same order of magnitude as the NLO QCD correction to the operator
$O_{tW}$. In some sense this effect can be considered as a $\mathcal{O}\left(
\Lambda^{-2} \right)$ modification to the standard QCD correction, and will
shift the NLO correction to the operator $O_{tW}$.

On the other hand, $F_+$ is more interesting because, at order
$\mathcal{O}\left( \Lambda^{-2} \right)$ and in the $m_b=0$ limit, $F_+$
vanishes at tree level even if anomalous $tbW$ couplings are present.  The
contributions to $F_+$ can only come from either a non-zero $m_b$ effect or a
$\mathcal{O}(\alpha_s)$ real gluon emission.  For this reason the contribution
from $O_{tG}$ is relatively more important.  In addition, the LHC sensitivity
on $F_+$ is better than the other two helicities, at 2 per-mille level for
$L=10\mathrm{fb}^{-1}$ \cite{AguilarSaavedra:2007rs}. Our numerical result
shows that the contribution from $O_{tG}$ is about 1/3 of that from $O_{tW}$,
and is not a negligible effect. Note the $-0.04C_{tW}$ in the first line of
Eq.~(\ref{eq:Fplus}) is a finite $m_b$ effect.

\begin{figure*}[tb]
  \begin{center}
    \includegraphics[width=0.9\linewidth]{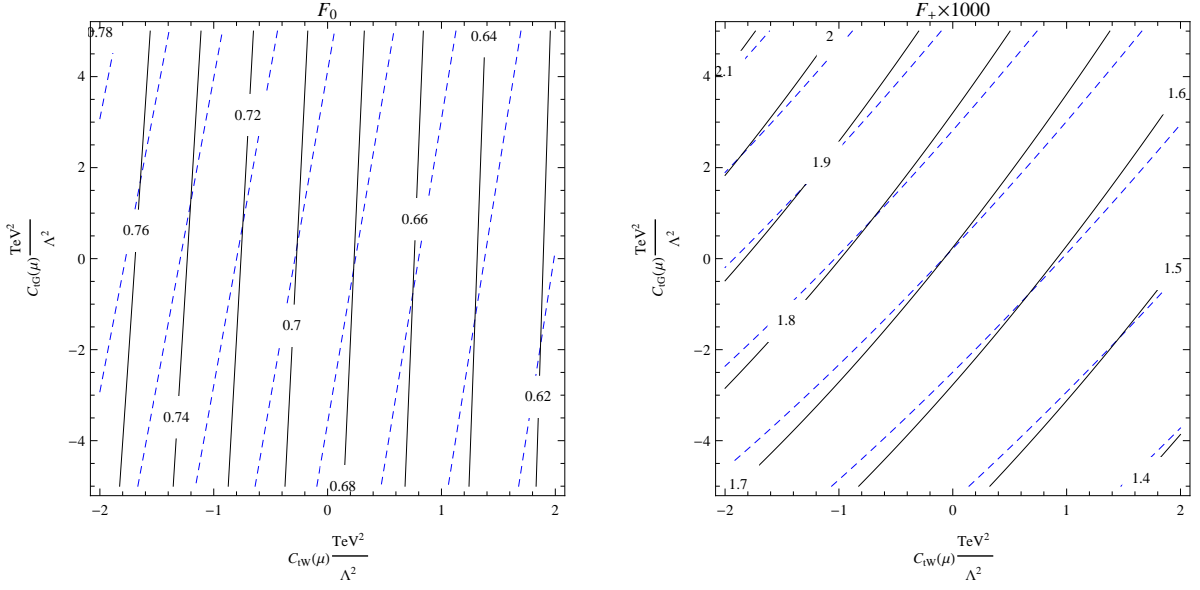}
  \end{center}
  \caption{Contour plots for $W$ helicity fractions
    in the $C_{tW}(\mu)$ - $C_{tG}(\mu)$ plane.  Left: $F_0$.  Right:
    $1000\times F_+$.
    The solid (black) curves represent $\mu=m_t$, while
  the dashed (blue) ones are for $\mu=2$ TeV.}
  \label{fig:tWtG}
\end{figure*}

In the above results we have chosen $\mu=m_t$ as our renormalization scale, and
thus the operator coefficients in these results should be interpreted as being
defined at a low energy scale $m_t$. These coefficients are probed by low
energy experiments, but they are not directly related to the new physics, which
resides at a higher scale $\Lambda$. In general, by matching the new physics to
an effective field theory at scale $\Lambda$, one obtains coefficients of
operators, $C_i(\Lambda)$, defined at scale $\Lambda$, and they need to be
evolved down to the scale $m_t$, to be compared with experimental results.  The
operator evolution of $O_{tW}$ and $O_{tG}$ is given in
Section~\ref{sec:mixings}.  In some cases one might be more interested in
$C_i(\Lambda)$ rather than $C_i(m_t)$ because they are directly related to the
new physics.  For this reason it is also useful to present results in terms of
$C_{tW,tG}(\Lambda)$, taking into account the running and mixing of the
coefficients. The results then depend on the scale $\Lambda$.  Assuming
$\Lambda=2$ TeV, the helicity fractions are
\begin{flalign}
  F_0=&0.689-0.038C_{tW}(\Lambda)\left(\frac{1\mathrm{TeV}}{\Lambda} \right)^2
  \nonumber\\&
  +0.0026C_{tG}(\Lambda)\left(\frac{1\mathrm{TeV}}{\Lambda} \right)^2
  \nonumber\\
  F_+=&\left[1.69-0.100C_{tW}(\Lambda)\left(\frac{1\mathrm{TeV}}{\Lambda} \right)^2
    \right.  \nonumber\\&\left.
    +0.038C_{tG}(\Lambda)\left(\frac{1\mathrm{TeV}}{\Lambda} \right)^2\right]
    \times 10^{-3}
    \ .
\end{flalign}
We can see the contribution of $O_{tG}$ in $F_0$ is enhanced by a factor of
$\sim4$ due to its mixing into $O_{tW}$.  To compare the results at scale
$\mu=m_t$ and scale $\mu=\Lambda$, we show the contour plots for $F_0$ and $F_+$
at two different scales in Figure~\ref{fig:tWtG}.

\subsection{$t\to bl\nu$}

In this section we discuss the effects of four-fermion operators on
$W$-helicity fractions.  The $W$-helicity fractions have been measured by both
ATLAS and CMS.  The results are normally used to set limits on the anomalous
$tbW$ couplings, $V_{L,R}$ and $g_{L,R}$, or alternatively on the coefficient
of the operator $O_{tW}$ (or $O_{uW}^{33}$).  However, one can imagine that a
four-fermion contact operator involving $tbl\nu$ may also lead to decay with
the same $bl\nu$ final state, giving rise to a shift of the measured
``$W$-helicity fractions'', even though a real $W$ is not involved in the
process.  One example is that some new heavy particle, $W'$, will mediate the
decay through $t\to bW^{\prime *}\to bl\nu$. If the mass of $W'$ is much larger
than $m_t$, this process is well described by the four-fermion operator
$O_{lQ}^{(3)}$.

The contribution from four-fermion operators are typically small, because their
interference with the SM amplitude is suppressed by a small numerical factor
due to a cancellation in total rate between two phase space regions,
$m_{l\nu}<m_W$ and $m_{l\nu}>m_W$ (where $m_{l\nu}$ is the invariant mass of
the lepton and neutrino) in which the $W$ propagator changes sign
\cite{AguilarSaavedra:2010zi}.  Their squared amplitude do not suffer from this
cancellation but are suppressed by an additional factor of $1/\Lambda^2$.
However, so far some of these operators can be probed only in the top decay
process, and one might hope to bound these operators using the differential
decay rate of $t\to bl\nu$. Furthermore, the invariant mass distribution of the
lepton pairs are also sensitive to the contact interaction, and probing the
off-shell region of $m_{l\nu}$ may provide some information about the
four-fermion operators.  Unfortunately, so far experimental measurements on
$W$-helicity fractions have required that $m_{l\nu}$ is equal to $m_W$, in
order to determine the momentum of the neutrino.  This condition certainly does
not apply to the case where the decay is mediated by a heavy $W'$, and so the
current limits on the helicity fractions cannot be used to put limits on
four-fermion operators.

The ``$W$-helicity fraction'' is extracted from the $\cos\theta$ distribution
of the decay rate, and in principle this can be done even if a real $W$ boson
is not involved, \ie when a four-fermion operator is present.  It is well known
that the SM and its interference with two-fermion operator $O_{\varphi
Q}^{(3)}$ produces $W$-helicities with the following ratio (at leading order):
\begin{equation}
  F_+:F_0:F_-=0:1:2x^2
  \ ,
\end{equation}
where $x=m_W/m_t$,
while its interference with two-fermion operator $O_{tW}$ gives:
\begin{equation}
  F_+:F_0:F_-=0:1:2
  \ ,
\end{equation}
note $F_+$ is always zero.  If the contribution comes from four-fermion
operators $O_{leQt}^{(1)}$ and $O_{lebQ}$, which take the form of a
scalar-current interaction, then the resulting differential decay rate has no
dependence on $\cos\theta$.  One would measure 
\begin{equation}
  F^+:F^0:F^-=1:1:1
\end{equation}
from the angular distribution.  On the other hand, the operator $O_{leQt}^{(3)}$
corresponds to a tensor-current interaction, and will give rise to different
helicity fractions with the following ratio:
\begin{equation}
  F^+:F^0:F^-=x^2:2-x^2:x^2
  \ ,
\end{equation}
where here $x=m_{l\nu}/m_t$ is not fixed at $m_W/m_t$.  The vector-current
operator $O_{lQ}^{(3)}$ on the other hand gives rise to a $V-A$ interaction,
and so its contribution to the helicity fractions has the same ratio as in the
SM, \ie 
\begin{equation}
  F^+:F^0:F^-=0:1:2x^2
  \ .
\end{equation}
Nevertheless the interference between this operator and the SM will cause a
shift in the distribution of $m_{l\nu}$, which in turn will modify the helicity
fractions.

In Table~\ref{tab:helicity} we present the numerical results for the helicity
fractions, from the SM contribution squared, the interference between the SM
and the two-fermion operators $O_{\varphi Q}^{(3)}$ and $O_{tW}$ as well as the
four-fermion operator $O_{lQ}^{(3)}$, and the squared contribution from
four-fermion operators.  We have assumed $\Lambda=1$ TeV and all coefficients
are equal to one.  To present the results in a more useful way, we divide the
phase space region into three sub-regions: 15 GeV$<m_{l\nu}<$70.4 GeV, 70.4
GeV$<m_{l\nu}<$90.4 GeV, and $m_{l\nu}>$90.4 GeV.  They are chosen such that
the second region corresponds to essentially two-body decay $t\to bW$ and
incorporates most contributions from two-fermion operators.  The results given
in the table are obtained by integrating over each phase space sub-region.  The
NLO QCD corrections, the finite $m_b$ effect at LO, and the finite $W$-width
corrections are all taken into account. The magnitudes of NLO corrections are
also given in the table.  One can see that in the second sub-region, \ie the
``on-shell'' region, the contribution is dominated by the SM and its
interference with two-fermion operators, and therefore helicity fractions
measured in this region provide information on $tbW$ couplings.  On the other
hand, in the other two ``off-shell'' regions the SM and two-fermion operators
are suppressed, and four-fermion contributions at order $\Lambda^{-4}$ are
comparable with the two-fermion contributions at order $\Lambda^{-2}$.  In
particular the four-fermion operators will modify the positive helicity
fraction which is almost zero in the SM, and this may help to determine or to
constrain the coefficient of four-fermion operators.

  \begin{table*}
    \[
\begin{array}{c|ccccc|ccccc|ccccc}
  &\multicolumn{5}{c|}{\text{Region 1:}\ [15,70.4]\ \text{GeV}}
  &\multicolumn{5}{c|}{\text{Region 2:}\ [70.4,90.4]\ \text{GeV}}
  &\multicolumn{5}{c}{\text{Region 3:}\ [90.4,\infty]\ \text{GeV}} \\ 
 \text{} & \text{Total(GeV)} & \text{NLO} & F_+ & F_0 & F_- &
   \text{Total(GeV)} & \text{NLO} & F_+ & F_0 & F_- & \text{Total(GeV)}
   & \text{NLO} & F_+ & F_0 & F_- \\\hline
 \text{SM}^2 & 4.2\times 10^{-3} & -9\% & 0.00 & 0.80 & 0.20 &
   1.4\times 10^{-1} & -9\% & 0.00 & 0.69 & 0.31 & 3.1\times 10^{-3} &
   -9\% & 0.00 & 0.58 & 0.42 \\
 \text{SM}\times C_{\varphi Q}^{(3)} & 5.1\times 10^{-4} &
   -9\% & 0.00 & 0.80 & 0.20 & 1.7\times 10^{-2} & -9\% & 0.00 & 0.69
   & 0.31 & 3.8\times 10^{-4} & -9\% & 0.00 & 0.58 & 0.42 \\
 \text{SM}\times C_{tW} & 3.\times 10^{-4} & -8\% & 0.00 & 0.34 &
   0.66 & 1.5\times 10^{-2} & -8\% & 0.00 & 0.34 & 0.66 & 4.8\times
   10^{-4} & -8\% & 0.00 & 0.34 & 0.66 \\
 \text{SM}\times C_{lQ}^{(3)} & -4.6\times 10^{-4} & -9\% &
   0.00 & 0.83 & 0.17 & 2.8\times 10^{-6} & 1\% & 0.05 & -8.75 & 9.70
   & 4.2\times 10^{-4} & -9\% & 0.00 & 0.54 & 0.46 \\
 C_{lQ}^{(3)2} & 1.4\times 10^{-5} & -9\% & 0.00 & 0.85
   & 0.14 & 8.7\times 10^{-6} & -9\% & 0.00 & 0.69 & 0.31 & 1.9\times
   10^{-5} & -9\% & 0.01 & 0.46 & 0.53 \\
 C_{leQt}^{(1)2} & 1.1\times 10^{-7} & -0\% & 0.33 &
   0.33 & 0.33 & 1.4\times 10^{-7} & 0\% & 0.33 & 0.33 & 0.33 &
   5.8\times 10^{-7} & 2\% & 0.33 & 0.33 & 0.33 \\
 C_{lebQ}^2 & 1.1\times 10^{-7} & -0\% & 0.33 & 0.33 & 0.33 &
   1.4\times 10^{-7} & 0\% & 0.33 & 0.33 & 0.33 & 5.8\times 10^{-7} &
   2\% & 0.33 & 0.33 & 0.33 \\
 C_{leQt}^{(3)2} & 1.3\times 10^{-5} & -9\% & 0.04 &
   0.92 & 0.04 & 6.9\times 10^{-6} & -9\% & 0.10 & 0.79 & 0.10 &
   1.5\times 10^{-5} & -8\% & 0.19 & 0.62 & 0.19 \\
\end{array}
    \]
      \caption{Numerical values of the helicity fractions at NLO, from
	the SM contribution, its interference with $O_{\varphi Q}^{(3)}$,
	$O_{tW}$ and $O_{lQ}^{(3)}$, and the squared contributions of
	four-fermion operators. The phase space is divided in to three regions
	according to $m_{l\nu}$, in such a way that the second sub-region is
	near the $W$ shell and incorporates almost all ``on shell'' decays.  For
	each region we present the total width, the NLO correction
	($\Gamma_{\rm NLO}/\Gamma_{\rm LO}-1$), and the three helicity
      fractions.}
	\label{tab:helicity}
  \end{table*}

\subsection{$t\to ul^+l^-$}

For flavor-changing decays $t\to ul^+l^-$, the situation is very similar.
Searches for $t\to uZ$ has been performed by assuming the lepton pair comes
from an on-shell $Z$, and cuts on invariant mass of the leptons near the
$Z$-mass shell are applied.  In principle, one should consider also the
possibility that FCNC is mediated by new heavy particles like $Z'$.  In this
case the decay process is described by four-fermion contact interactions, and
so one should take into account their squared contributions as well as their
interferences with $t\to u\gamma^*,uZ^*\to ul^+l^-$.  Unlike the main decay
channel where the interference between SM and four-fermion operators is
suppressed, in the flavor-changing decay mode all contributions start at order
$\Lambda^{-4}$, so the four-fermion operators are relatively more important.

To present the decay rate from each operator, we write the decay width of $t\to
ue^+e^-$ as
\begin{flalign}
  \Gamma_{t\to ue^+e^-}=\sum_{i,j}\Gamma_{ij}C_i(m_t)C_j(m_t)\frac{(1\ \rm
  TeV)^2}{\Lambda^4}
  \ ,
  \label{eq:numgamma}
\end{flalign}
where $C_i(m_t)$ are either the real or the imaginary part of the operator
coefficients, defined at scale $m_t$.  In Table~\ref{tab:fcnc0} we show the
numerical values of $\Gamma_{ij}$ for some typical operators, including
$O_{\varphi q}^{(1,1+3)}$ which gives rise to the vector coupling of $Z$,
$O_{uW,uG}^{(13)}$ which gives rise to the tensor couplings of $Z$, $\gamma$
and $g$, and the S-S, V-V and T-T types four-fermion operators, assuming their
coefficients are real.  These numerical results are obtained with a cut on the
invariant mass of the lepton pair, $m_{ll}>15$~GeV. This is required not only
to remove not only the QCD background but also the divergence from the photon
peak, which comes from the weak-dipole operators $O_{uB,uW}^{(13),(31)}$.  In
the table we also show the amount of NLO corrections. The complete results for
all operators are given in Appendix~\ref{sec:apptable}.  Note that the
interference between S-S and T-T operators is proportional to $\cos\theta$, and
vanishes only after integrating over the $\theta$ angle.  The QCD correction to
this part is about $-2\%$.

If the flavor-changing decays are observed at the LHC, the next step will be to
determine the specific form of the flavor-changing interactions.  To this end
one needs to make use of the kinematic information of the final state leptons.
We have provided results for the differential decay rates, including the
invariant-mass distribution and the angular distribution of the final state
leptons, and we expect these results will be useful in future analyses for
flavor-changing top-quark interactions.

\begin{table*}
  \begin{minipage}[t]{.68\linewidth}
    \[
\begin{array}{c|cccc}
  \Gamma_{ij} (\mathrm{GeV}) & \Re\left(C_{\varphi q}^{(1,1+3)}\right) &
  \Re\left(C_{uW}^{(13)}\right) & \Re\left(C_{uG}^{(13)}\right) &
  \Re\left(C_{lq}^{(1,1+3)}\right) \\\hline
 \Re\left(C_{\varphi q}^{(1,1+3)}\right) & \underset{-8\%}{1.9\times 10^{-5}} &
 \underset{-8\%}{-6.2\times 10^{-5}} & \underset{\text{---}}{2.9 \times 10^{-6}} & \underset{-12\%}{-3.5\times 10^{-7}} \\
 \Re\left(C_{uW}^{(13)}\right) & \underset{}{} & \underset{-9\%}{7.6\times 10^{-5}} & \underset{\text{---}}{-6.1\times 10^{-6}} & \underset{-7\%}{-3.3\times 10^{-6}} \\
 \Re\left(C_{uG}^{(13)}\right) & \underset{}{} & \underset{}{} & \underset{\text{---}}{6.8\times 10^{-8}} & \underset{\text{---}}{2.6\times 10^{-7}} \\
 \Re\left(C_{lq}^{(1,1+3)}\right) & \underset{}{} & \underset{}{} & \underset{}{} & \underset{-8\%}{2.9\times 10^{-6}} \\
\end{array}
    \]
  \end{minipage}
  \begin{minipage}[t]{.3\linewidth}
    \[
\begin{array}{c|cc}
 \Gamma_{ij} (\mathrm{GeV}) & \Re\left(C_{lequ}^{(1,13)}\right) &
 \Re\left(C_{lequ}^{(3,13)}\right) \\\hline
 \Re\left(C_{lequ}^{(1,13)}\right) & \underset{1\%}{8.2\times 10^{-7}} &
 \underset{\text{---}}{0.} \\
 \Re\left(C_{lequ}^{(3,13)}\right) & \underset{}{} & \underset{-8\%}{3.5\times 10^{-5}} \\
\end{array}
      \]
  \end{minipage}
  \caption{\label{tab:fcnc0}
  Numerical values for $\Gamma_{ij}$ from some typical operators.  Left:
  Two-fermion operators and V-V type four-fermion operator, and their
  interferences.  Right: S-S and T-T type four-fermion operators and their
  interferences.  The percentage number under each entry represents the amount
  of NLO correction (a dash implies the presented value vanishes at LO).}
\end{table*}

\section{Summary}
\label{sec:conclusion}
Measurements on top-quark related processes can provide valuable information on
new physics.  In general, non-standard interactions of the top quark should be
studied in a global manner, in particular due to their mixing effects at NLO in
QCD.  For this reason a model-independent analysis based on an effective field
theory approach should be performed, and a global fit needs to be done,
including all available measurements and all dimension-six operators.

In this paper we have presented the complete calculation for top-quark
semi-leptonic decays in the presence of new physics, at the NLO accuracy in QCD
in an EFT approach. We have employed the operator basis in
Ref.~\cite{Grzadkowski:2010es}, and consider all dimension-six operators that
give rise to a non-standard interaction of the top quark and contribute to
top-quark decay processes, including both flavor-conserving and flavor-changing
decay modes.  Apart from confirming results that were previously available in
the literature, we have taken into account the QCD corrections from the
color-dipole operators, the differential decay rate of semi-leptonic final
states, and the contributions from four-fermion operators.  

Our results are presented in terms of analytical expressions for total and
differential decay rates as well as their numerical evaluation.  The QCD
corrections can reach the ten-percent level depending on the processes and
operators. In addition, in many cases new contributions enter at the NLO,
\textit{e.g.} from color-dipole operators.  For completeness we have also
presented the $\mathcal{O}(\alpha_s)$ mixing of all relevant operators.

Our results complete the set of calculations needed for a model-independent
study of top-quark decay at NLO accuracy and therefore provide all information
needed to perform global analyses for new physics searches.

\section{Acknowledgements}
I am grateful for valuable discussions with Gauthier Durieux and Fabio Maltoni. 
The work is supported by the IISN ``Fundamental
interactions'' convention 4.4517.08.

\newpage

\appendix

\section{NLO corrections}
\label{sec:appNLO}
\newcommand{\nl}{\nonumber\\&}
\newcommand{\nll}{\right.\nonumber\\&\qquad\left.}
\newcommand{\nlll}{\right.\right.\nonumber\\&\qquad\qquad\left.\left.}
\begin{widetext}
In the following we collect all $\mathcal{O}(\alpha_s)$ corrections that
appeared in the main text.

The $\delta^{i}_{V,VT,T,VG}(x)$ functions, appeared in
Eq.~(\ref{eq:fulltbw1}-\ref{eq:fulltbw2}) and Eq.~(\ref{eq:fulltuz2}), are:
\begin{flalign}
  \delta_{V}^{\rm tot}(x)&=
-\frac{1}{9 \pi  \left(x^2-1\right)^2 \left(2 x^2+1\right)}
\Bigg\{
+\left(x^2-1\right) \left(3 \left(-6 x^4+9 x^2+5\right)+\pi ^2 \left(8 x^4-4 x^2-4\right)\right)
\nl
+6 \left(x^2-1\right)^2 \left(4 x^2+5\right) \log (1-x)
-24 x^2 \left(2 x^4+x^2-1\right) \log (x)
+24 \left(2 x^6-3 x^4+1\right) \log (1-x) \log (x)
\nl
+6 \left(x^2-1\right)^2 \left(4 x^2+5\right) \log (x+1)
+24 \left(2 x^6-3 x^4+1\right) \log (x) \log (x+1)
+48 \left(2 x^6-3 x^4+1\right) \text{Li}_2(-x)
\nl
+48 \left(2 x^6-3 x^4+1\right) \text{Li}_2(x)
\Bigg\}
\end{flalign}
\begin{flalign}
  \delta_{VT}^{\rm tot}(x)&=
-\frac{1}{9 \pi  x^2 \left(x^2-1\right)^2}
  \Bigg\{
-3 x^2 \left(x^2-1\right)^2 \log \left(\frac{m_t^2}{\mu ^2}\right)
+x^2 \left(x^2-1\right) \left(-21 x^2+4 \pi ^2 \left(x^2-1\right)+17\right)
\nl
+2 \left(x^2-1\right)^2 \left(7 x^2+2\right) \log (1-x)
+8 x^4 \left(3-2 x^2\right) \log (x)
+24 x^2 \left(x^2-1\right)^2 \log (1-x) \log (x)
\nl
+2 \left(x^2-1\right)^2 \left(7 x^2+2\right) \log (x+1)
+24 x^2 \left(x^2-1\right)^2 \log (x) \log (x+1)
+48 x^2 \left(x^2-1\right)^2 \text{Li}_2(-x)
\nl
+48 x^2 \left(x^2-1\right)^2 \text{Li}_2(x)
\Bigg\}
\end{flalign}
\begin{flalign}
  \delta_{T}^{\rm tot}(x)&=
-\frac{1}{9 \pi  \left(x^2-1\right)^2 \left(x^2+2\right)}
\Bigg\{
-6 \left(x^6-3 x^2+2\right) \log \left(\frac{m_t^2}{\mu ^2}\right)
\nl
+\left(x^2-1\right) \left(-7 x^4-13 x^2+4 \pi ^2 \left(x^4+x^2-2\right)+32\right)
+6 \left(x^2-1\right)^2 \left(x^2+8\right) \log (1-x)
\nl
-24 x^2 \left(x^4+2 x^2-2\right) \log (x)
+24 \left(x^6-3 x^2+2\right) \log (1-x) \log (x)
+6 \left(x^2-1\right)^2 \left(x^2+8\right) \log (x+1)
\nl
+24 \left(x^6-3 x^2+2\right) \log (x) \log (x+1)
+48 \left(x^6-3 x^2+2\right) \text{Li}_2(-x)
+48 \left(x^6-3 x^2+2\right) \text{Li}_2(x)
\Bigg\}
\end{flalign}
\begin{flalign}
  \delta_{VG}^{\rm tot}(x)&=
\frac{1}{18 \pi  x^2 \left(x^2-1\right)^2}
\Bigg\{
+6 x^2 \left(x^2-1\right)^2 \log \left(\frac{m_t^2}{\mu^2}\right)
-1 -10 x^2 +31 x^4 -20 x^6
+4 \left(x^2-1\right)^3 \log (1-x)
\nl
+4 x^4 \left(x^2+3\right) \log (x)
+4 \left(x^2-1\right)^3 \log (x+1)
\Bigg\}
\end{flalign}
\begin{flalign}
  \delta_{V}^{L}(x)&=
-\frac{1}{9 \pi  \left(x^2-1\right)^2}
\Bigg\{
+4 \pi ^2 \left(2 x^4+5 x^2+1\right)
-3 \left(4 x^6-51 x^4+42 x^2+5\right)
+18 \left(x^2-1\right)^2 \log (1-x)
\nl
-96 \left(2 x^4+x^2\right) \log (x)
+12 (x-1)^2 \left(x^3+6 x^2-x+2\right) \log (1-x) \log (x)
+18 \left(x^2-1\right)^2 \log (x+1)
\nl
-12 (x+1)^2 \left(x^3-6 x^2-x-2\right) \log (x) \log (x+1)
-12 (x+1)^2 \left(x^3-8 x^2+3 x-4\right) \text{Li}_2(-x)
\nl
+12 (x-1)^2 \left(x^3+8 x^2+3 x+4\right) \text{Li}_2(x)
\Bigg\}
\end{flalign}
\begin{flalign}
  \delta_{VT}^{L}(x)&=
-\frac{1}{9 \pi  x^2 \left(x^2-1\right)^2}
\Bigg\{
-3 x^2 \left(x^2-1\right)^2 \log \left(\frac{m_t^2}{\mu ^2}\right)
+x^2 \left(-33 x^4+30 x^2+2 \pi ^2 \left(2 x^4-7 x^2+1\right)+3\right)
\nl
+6 \left(2 x^6-3 x^4+1\right) \log (1-x)
-12 x^4 \left(x^2-7\right) \log (x)
+12 (x-1)^2 x^2 \left(2 x^2+5 x+1\right) \log (1-x) \log (x)
\nl
+6 \left(2 x^6-3 x^4+1\right) \log (x+1)
+12 x^2 (x+1)^2 \left(2 x^2-5 x+1\right) \log (x) \log (x+1)
\nl
+12 x^2 (x+1)^2 \left(4 x^2-9 x+3\right) \text{Li}_2(-x)
+12 (x-1)^2 x^2 \left(4 x^2+9 x+3\right) \text{Li}_2(x)
\Bigg\}
\end{flalign}
\begin{flalign}
  \delta_{T}^{L}(x)&=
-\frac{1}{9 \pi  x^2 \left(x^2-1\right)^2}
\Bigg\{
-6 x^2 \left(x^2-1\right)^2 \log \left(\frac{m_t^2}{\mu ^2}\right)
+x^2 \left(3 \left(x^4-22 x^2+21\right)+4 \pi ^2 \left(x^4-2 x^2-3\right)\right)
\nl
+6 \left(x^6-3 x^2+2\right) \log (1-x)
-24 x^2 \left(x^4-3 x^2-3\right) \log (x)
+12 (x-1)^2 x \left(2 x^3+3 x^2+3\right) \log (1-x) \log (x)
\nl
+6 \left(x^6-3 x^2+2\right) \log (x+1)
+12 x (x+1)^2 \left(2 x^3-3 x^2-3\right) \log (x) \log (x+1)
\nl
+12 x (x+1)^2 \left(4 x^3-7 x^2+2 x-3\right) \text{Li}_2(-x)
+12 (x-1)^2 x \left(4 x^3+7 x^2+2 x+3\right) \text{Li}_2(x)
\Bigg\}
\end{flalign}
\begin{flalign}
  \delta_{VG}^{L}(x)&=
\frac{1}{18 \pi  x^2 \left(x^2-1\right)^2}
\Bigg\{
+6 x^2 \left(x^2-1\right)^2 \log \left(\frac{m_t^2}{\mu^2}\right)
-3 +2 \left(\pi ^2-12\right) x^2
+\left(33+6 \pi ^2\right) x^4
-6 x^6
\nl
+6 \left(x^2-1\right)^3 \log (1-x)
-48 x^4 \log (x)
-12 (x-1)^3 x^2 \log (1-x) \log (x)
+6 \left(x^2-1\right)^3 \log (x+1)
\nl
+12 x^2 (x+1)^3 \log (x) \log (x+1)
+12 x^2 (x+1)^3 \text{Li}_2(-x)
-12 (x-1)^3 x^2 \text{Li}_2(x)
\Bigg\}
\end{flalign}
\begin{flalign}
  \delta_{V}^{T}(x)&=
-\frac{1}{9 \pi  x^2 \left(x^2-1\right)^2}
\Bigg\{
+x^2 \left(2 \pi ^2 \left(2 x^4-5 x^2-5\right)-3 \left(x^4+18 x^2-19\right)\right)
+6 \left(2 x^6-3 x^4+1\right) \log (1-x)
\nl
+12 x^2 \left(-2 x^4+7 x^2+5\right) \log (x)
+6 (x-1)^2 x \left(4 x^3+7 x^2+5\right) \log (1-x) \log (x)
+6 \left(2 x^6-3 x^4+1\right) \log (x+1)
\nl
+6 x (x+1)^2 \left(4 x^3-7 x^2-5\right) \log (x) \log (x+1)
+6 x (x+1)^2 \left(8 x^3-15 x^2+4 x-5\right) \text{Li}_2(-x)
\nl
+6 (x-1)^2 x \left(8 x^3+15 x^2+4 x+5\right) \text{Li}_2(x)
\Bigg\}
\end{flalign}
\begin{flalign}
  \delta_{VT}^{T}(x)&=
-\frac{1}{9 \pi  x^2 \left(x^2-1\right)^2}
\Bigg\{
-3 x^2 \left(x^2-1\right)^2 \log \left(\frac{m_t^2}{\mu ^2}\right)
+x^2 \left(\pi ^2 \left(4 x^4-5 x^2+5\right)-3 \left(5 x^4-14 x^2+9\right)\right)
\nl
+3 \left(x^2-1\right)^2 \left(5 x^2+1\right) \log (1-x)
-6 \left(3 x^6+x^4\right) \log (x)
+6 (x-1)^2 x^2 \left(4 x^2+7 x+5\right) \log (1-x) \log (x)
\nl
+3 \left(x^2-1\right)^2 \left(5 x^2+1\right) \log (x+1)
+6 x^2 (x+1)^2 \left(4 x^2-7 x+5\right) \log (x) \log (x+1)
\nl
+6 x^2 (x+1)^2 \left(8 x^2-15 x+9\right) \text{Li}_2(-x)
+6 (x-1)^2 x^2 \left(8 x^2+15 x+9\right) \text{Li}_2(x)
\Bigg\}
\end{flalign}
\begin{flalign}
  \delta_{T}^{T}(x)&=
-\frac{1}{9 \pi  \left(x^2-1\right)^2}
\Bigg\{
-6 \left(x^2-1\right)^2 \log \left(\frac{m_t^2}{\mu ^2}\right)
-16
-9 x^2
+30 x^4
-5 x^6
+4 \pi ^2 \left(x^4+1\right)
\nl
+18 \left(x^2-1\right)^2 \log (1-x)
-12 \left(5 x^4+x^2\right) \log (x)
+6 (x-1)^2 \left(x^3+6 x^2+5 x+4\right) \log (1-x) \log (x)
\nl
+18 \left(x^2-1\right)^2 \log (x+1)
+6 \left(-x^5+4 x^4+6 x^3+3 x+4\right) \log (x) \log (x+1)
\nl
-6 (x+1)^2 \left(x^3-10 x^2+13 x-8\right) \text{Li}_2(-x)
+6 (x-1)^2 \left(x^3+10 x^2+13 x+8\right) \text{Li}_2(x)
\Bigg\}
\end{flalign}
\begin{flalign}
  \delta_{VG}^{T}(x)&=
-\frac{1}{18 \pi  x^2 \left(x^2-1\right)^2}
\Bigg\{
-6 x^2 \left(x^2-1\right)^2 \log \left(\frac{m_t^2}{\mu^2}\right)
+x^2 \left(27 x^4-30 x^2+\pi ^2 \left(3 x^2+1\right)+3\right)
\nl
-3 \left(x^2-1\right)^3 \log (1-x)
-6 x^4 \left(x^2+7\right) \log (x)
-6 (x-1)^3 x^2 \log (1-x) \log (x)
-3 \left(x^2-1\right)^3 \log (x+1)
\nl
+6 x^2 (x+1)^3 \log (x) \log (x+1)
+6 x^2 (x+1)^3 \text{Li}_2(-x)
-6 (x-1)^3 x^2 \text{Li}_2(x)
\Bigg\}
\end{flalign}
\begin{flalign}
  \delta_{V}^{F}(x)&=
-\frac{2}{9 \pi  x^2 \left(x^2-1\right)^2}
\Bigg\{
+x^2 \left(\pi ^2 \left(x^2+2\right)+3 (4 x-3) (x-1)^2\right)
+\left(6 x^6-9 x^4+3\right) \log (1-x)
\nl
+\left(-6 x^6+33 x^4-30 x^2+3\right) \log (x+1)
-12 x^2 \left(x^4-3 x^2-1\right) \text{Li}_2(-x)
+12 x^2 \left(x^2-1\right)^2 \text{Li}_2(x)
\Bigg\}
\end{flalign}
\begin{flalign}
  \delta_{VT}^{F}(x)&=
-\frac{1}{9 \pi  x^2 \left(x^2-1\right)^2}
\Bigg\{
-3 x^2 \left(x^2-1\right)^2 \log \left(\frac{m_t^2}{\mu ^2}\right)
+x^2 \left(5 \pi ^2 \left(x^2+1\right)-3 (x-1)^2 \left(5 x^2-12 x+9\right)\right)
\nl
+3 \left(x^2-1\right)^2 \left(5 x^2+1\right) \log (1-x)
+\left(-3 x^6+69 x^4-69 x^2+3\right) \log (x+1)
\nl
+12 x^2 \left(-2 x^4+9 x^2+3\right) \text{Li}_2(-x)
+24 x^2 \left(x^2-1\right)^2 \text{Li}_2(x)
\Bigg\}
\end{flalign}
\begin{flalign}
  \delta_{T}^{F}(x)&=
\frac{1}{9 \pi  \left(x^2-1\right)^2}
\Bigg\{
+6 \left(x^2-1\right)^2 \log \left(\frac{m_t^2}{\mu ^2}\right)
+2 \pi ^2 \left(x^4-2\right)
+(x-1)^2 \left(x^4+2 x^3-9 x^2-28 x+16\right)
\nl
-18 \left(x^2-1\right)^2 \log (1-x)
+6 \left(5 x^4-12 x^2+7\right) \log (x+1)
+24 \left(2 x^4-2 x^2-1\right) \text{Li}_2(-x)
-24 \left(x^2-1\right)^2 \text{Li}_2(x)
\Bigg\}
\end{flalign}
\begin{flalign}
  \delta_{VG}^{F}(x)&=
\frac{1}{18 \pi  x^2 \left(x^2-1\right)^2}
\Bigg\{
+6 x^2 \left(x^2-1\right)^2 \log \left(\frac{m_t^2}{\mu^2}\right)
+x^2 \left(\pi ^2 \left(3 x^2+1\right)-6 (x-1)^2 \left(5 x^2+3 x+4\right)\right)
\nl
+3 \left(x^2-1\right)^3 \log (1-x)
+3 \left(3 x^6+x^4-3 x^2-1\right) \log (x+1)
+12 \left(3 x^4+x^2\right) \text{Li}_2(-x)
\Bigg\}
\ .
\end{flalign}

In the following we define
\begin{equation}
  y\equiv\frac{x+i\sqrt{4-x^2}}{2}
  \ .
\end{equation}
The $\delta^{i}_{VG,TG,G2}(x)$ functions that appeared in Eq.~(\ref{eq:fulltuz2})
are:
\begin{flalign}
  \delta_{VG,r}^{tot}(x,v,a)&=
\frac{1}{54 \pi  x^2 \left(x^2-1\right)^2}
\Bigg\{
+36 v x^2 \left(x^2-1\right)^2 \log \left(\frac{m_t^2}{\mu^2}\right)
+v x^2 \left(-81 x^4+156 x^2-6 \pi  \sqrt{4-x^2} x^3+4 \pi ^2-75\right)
\nl
+a \left(-39 x^6+66 x^4+\left(4 \pi ^2-21\right) x^2-6 \pi  \sqrt{4-x^2} x^5+18 \pi  \sqrt{4-x^2} x^3-6\right)
\nl
+12 x \sin ^{-1}\left(\frac{x}{2}\right) \left(a \left(\sqrt{4-x^2} \left(x^4-3 x^2-4\right)-2 \pi  x\right)+v \left(\sqrt{4-x^2} \left(7 x^4-6 x^2+2\right)-2 \pi  x\right)\right)
\nl
-24 x^2 (a+v) \sin ^{-1}\left(\frac{x}{2}\right)^2
-12 x^2 \log (x) \left(a \left(x^4-5 x^2-3\right)+v \left(x^4+4 x^2-3\right)\right)
\nl
+24 x^2 (a+v) \Re\text{Li}_2(x y)
\Bigg\}
\end{flalign}
\begin{flalign}
  \delta_{VG,i}^{tot}(x,v,a)&=-\frac{1}{3}(v+a)
\end{flalign}
\begin{flalign}
  \delta_{TG,r}^{tot}(x,v,a)&=
\frac{1}{9 \pi  x \left(x^2-1\right)^2 \left(x^2+2\right)}
\Bigg\{
+6 v x \left(x^6-3 x^2+2\right) \log \left(\frac{m_t^2}{\mu^2}\right)
\nl
+x \left[a \left(-3 \pi  x \sqrt{4-x^2} \left(x^2-2\right)+3 \left(x^4+14 x^2-15\right)+4 \pi ^2\right)
\nll
+v \left(-11 x^6-12 x^4+81 x^2-6 \pi  \sqrt{4-x^2} \left(x^2-1\right) x+4 \pi ^2-58\right)\right]
\nl
-6 \sin ^{-1}\left(\frac{x}{2}\right) \left(a \left(\sqrt{4-x^2} \left(x^6-5 x^4+3 x^2+4\right)+4 \pi  x\right)+v x \left(4 \pi -x \sqrt{4-x^2} \left(x^4+8 x^2-9\right)\right)\right)
\nl
-24 x (a+v) \sin ^{-1}\left(\frac{x}{2}\right)^2
+6 x^3 \log (x) \left(a \left(x^4-7 x^2-7\right)+v \left(x^4-6 x^2-3\right)\right)
+24 x (a+v) \Re\text{Li}_2(x y)
\Bigg\}
\end{flalign}
\begin{flalign}
  \delta_{TG,i}^{tot}(x,v,a)&=-\frac{1}{3}(v+a)
\end{flalign}
\begin{flalign}
  \delta_{G2}^{tot}(x,v,a)&=
-\frac{1}{36 \pi  x^2 \left(x^2-4\right) \left(x^2-1\right)^2 \left(x^2+2\right)}
\Bigg\{
-a^2 \left(x^2-4\right) \left(5 x^8-43 x^6+108 x^4-\left(71+2 \pi ^2\right) x^2
\nll
-9 \pi  \sqrt{4-x^2} x^5+60 \pi  \sqrt{4-x^2} x^3+1\right)
-2 a v x^2 \left(x^2-4\right) \left(45 x^4-66 x^2+3 \pi  \sqrt{4-x^2} \left(x^2+2\right) x-2 \pi ^2+21\right)
\nl
+v^2 x^2 \left(-4 x^8+19 x^6+66 x^4-389 x^2+2 \pi ^2 \left(x^2-4\right)+3 \pi  \sqrt{4-x^2} \left(3 x^4-10 x^2-20\right) x+308\right)
\nl
-6 x^2 \sin ^{-1}\left(\frac{x}{2}\right) \left(a^2 \left(x^2-4\right) \left(3 x \sqrt{4-x^2} \left(3 x^2-20\right)+4 \pi \right)+2 a v \left(x^2-4\right) \left(4 \pi -3 x \sqrt{4-x^2} \left(x^2+2\right)\right)
\nll
+v^2 \left(4 \pi  \left(x^2-4\right)+3 x \sqrt{4-x^2} \left(3 x^4-10 x^2-20\right)\right)\right)
\nl
+72 x^2 \left(x^2-4\right) (a+v)^2 \sin ^{-1}\left(\frac{x}{2}\right)^2
+6 x^2 \left(x^2-4\right) \log (x) \left(a^2 \left(9 x^4-46 x^2-2\right)+2 a v \left(5 x^4-6\right)
\nll
+v^2 \left(9 x^4+4 x^2-10\right)\right)
-24 x^2 \left(x^2-4\right) (a+v)^2 \log ^2(x)
\Bigg\}
\end{flalign}
\begin{flalign}
  \delta_{VG,r}^{L}(x,v,a)&=
-\frac{1}{9 \pi  x^2 \left(x^2-1\right)^3}
\Bigg\{
-6 v x^2 \left(x^2-1\right)^3 \log \left(\frac{m_t^2}{\mu^2}\right)
+60 x^2 (a+v) \sin ^{-1}\left(\frac{x}{2}\right)^2
\nl
+a \left(3 x^8+3 \pi ^2 x^7-3 \left(3+2 \pi ^2\right) x^6+2 \left(3+\pi ^2\right) x^4-3 \pi ^2 x^3+3 \left(1+\pi ^2\right) x^2-3\right)
\nl
+v x^2 \left(3 \left(4 x^2-5\right) \left(x^2-1\right)^2+3 \pi  x \sqrt{4-x^2} \left(x^2-1\right)+\pi ^2 \left(3 x^5-6 x^4+2 x^2-3 x+3\right)\right)
\nl
-6 x \sqrt{4-x^2} \left(x^2-1\right) \sin ^{-1}\left(\frac{x}{2}\right) \left(v \left(x^4+2 x^2+2\right)-a x^2 \left(x^2-3\right)\right)
\nl
-6 x^2 \left(x^2-1\right) \log (x) \left(a \left(x^4-4 x^2+1\right)+v \left(x^4-7 x^2+1\right)\right)
\nl
-6 (x-1)^3 x \left(x^3+3 x^2+x-1\right) (a+v) \log (1-x) \log (x)
-12 x^4 \left(x^2-1\right) (a+v) \log ^2(x)
\nl
+6 x (x+1)^3 \left(x^3-3 x^2+x+1\right) (a+v) \log (x) \log (x+1)
-24 x^2 (a+v) \Re\text{Li}_2(x y)
\nl
+6 x (x+1)^3 \left(x^3-3 x^2+x+1\right) (a+v) \text{Li}_2(-x)
-6 (x-1)^3 x \left(x^3+3 x^2+x-1\right) (a+v) \text{Li}_2(x)
\Bigg\}
\end{flalign}
\begin{flalign}
  \delta_{VG,i}^{L}(x,v,a)&=-\frac{1}{3}(v+a)
\end{flalign}
\begin{flalign}
  \delta_{TG,r}^{L}(x,v,a)&=
-\frac{1}{9 \pi  x^3 \left(x^2-4\right) \left(x^2-1\right)^3}
\Bigg\{
-6 v x^3 \left(x^2-4\right) \left(x^2-1\right)^3 \log \left(\frac{m_t^2}{\mu^2}\right)
+24 x \left(3 v-a x^2 \left(x^2-4\right)\right) \Re\text{Li}_2(x y)
\nl
+x \left(a \left(x^2-4\right) \left(\left(2 x^4-40 x^2-31\right) \left(x^2-1\right)^2+\pi ^2 \left(2 x^6-12 x^5-4 x^4+12 x^3-3 x^2+4\right)\right)
\nll
+v x \left(12 x^9+\left(4 \pi ^2-69\right) x^7+\left(90-8 \pi ^2\right) x^5-3 \left(7+15 \pi ^2\right) x^3+3 \pi  \left(11 \sqrt{4-x^2}-16 \pi \right) x^2
\nlll
-18 \pi  \sqrt{4-x^2}+3 \pi  \left(\sqrt{4-x^2}-4 \pi \right) x^6+6 \pi  \left(10 \pi -3 \sqrt{4-x^2}\right) x^4+4 \left(13 \pi ^2-3\right) x\right)\right)
\nl
+6 x \left(x^2-1\right) \sin ^{-1}\left(\frac{x}{2}\right) \left(x \sqrt{4-x^2} \left(a x^2 \left(x^4-7 x^2+12\right)-v \left(x^6-2 x^4-9 x^2+10\right)\right)+4 \pi  v \left(x^2-3\right)\right)
\nl
+12 \sin ^{-1}\left(\frac{x}{2}\right)^2 \left(5 a x^3 \left(x^2-4\right)-3 v x \left(x^4-4 x^2+8\right)\right)
+12 x^3 \left(x^6-4 x^4-x^2+4\right) (a+v) \log ^2(x)
\nl
-6 x \left(x^4-5 x^2+4\right) \log (x) \left(a \left(x^6-8 x^4-15 x^2-2\right)+v \left(x^4-x^2+6\right) x^2\right)
\nl
-6 (x-1)^3 \left(x^2+3 x+2\right) \log (1-x) \log (x) \left(a \left(4 x^4-9 x^3+4 x^2-5 x+2\right)+v x \left(2 x^3-x^2-8 x+3\right)\right)
\nl
-6 (x+1)^3 \left(x^2-3 x+2\right) \log (x) \log (x+1) \left(a \left(4 x^4+9 x^3+4 x^2+5 x+2\right)+v x \left(2 x^3+x^2-8 x-3\right)\right)
\nl
-6 (x+1)^3 \left(x^2-3 x+2\right) \text{Li}_2(-x) \left(a \left(4 x^4+9 x^3+4 x^2+5 x+2\right)+v x \left(2 x^3+x^2-8 x-3\right)\right)
\nl
-6 (x-1)^3 \left(x^2+3 x+2\right) \text{Li}_2(x) \left(a \left(4 x^4-9 x^3+4 x^2-5 x+2\right)+v x \left(2 x^3-x^2-8 x+3\right)\right)
\Bigg\}
\end{flalign}
\begin{flalign}
  \delta_{TG,i}^{L}(x,v,a)&=-\frac{1}{3}(v+a)
\end{flalign}
\begin{flalign}
  \delta_{G2}^{L}(x,v,a)&=
-\frac{1}{36 \pi  (x-2) x^4 (x+2) \left(x^2-1\right)^2}
\Bigg\{
-a^2 \left(x^2-4\right) \left(x^8-28 x^6-15 \pi ^2 x^5+6 \left(\pi ^2-3\right) x^4-15 \pi ^2 x^3+44 x^2+1\right)
\nl
-v^2 x^2 \left(6 \pi  \left(10-3 x^2\right) \sqrt{4-x^2} x+12 \left(x^4-5 x^2+4\right) x^2+\pi ^2 \left(5 x^5-2 x^4+5 x^3+8 x^2-80 x+16\right)\right)
\nl
+2 a v x^2 \left(-6 \pi  x \sqrt{4-x^2} \left(x^2-6\right)+12 \left(x^6-7 x^4+14 x^2-8\right)+\pi ^2 \left(5 x^5-2 x^4-25 x^3+8 x^2+10 x-8\right)\right)
\nl
+12 v x^2 \sin ^{-1}\left(\frac{x}{2}\right) \left(a \left(6 x \sqrt{4-x^2} \left(x^2-6\right)+8 \pi \right)+v \left(3 x \sqrt{4-x^2} \left(10-3 x^2\right)+8 \pi \right)\right)
\nl
+96 v x^2 (a+v) \sin ^{-1}\left(\frac{x}{2}\right)^2
-12 x^2 \left(x^2-4\right) \log (x) \left(2 a^2 \left(5 x^2+1\right)+2 a v \left(5 x^2-2\right)-v^2 \left(7 x^2+2\right)\right)
\nl
-24 (x-1)^2 x^2 (x+2) \log (1-x) \log (x) \left(3 a^2 (x-2) x+2 a v \left(x^2-2 x-2\right)-v^2 \left(x^2-2 x+4\right)\right)
\nl
+6 x^2 (x+2) \log ^2(x) \left(3 a^2 x \left(x^3-8 x^2+13 x-2\right)
\nll
+2 a v \left(x^4-8 x^3+11 x^2+2 x-2\right)-v^2 \left(x^4-8 x^3+17 x^2-10 x+4\right)\right)
\nl
+24 (x-2) x^2 (x+1)^2 \log (x) \log (x+1) \left(3 a^2 x (x+2)+2 a v \left(x^2+2 x-2\right)-v^2 \left(x^2+2 x+4\right)\right)
\nl
+24 (x-2) x^2 (x+1)^2 \text{Li}_2(-x) \left(3 a^2 x (x+2)+2 a v \left(x^2+2 x-2\right)-v^2 \left(x^2+2 x+4\right)\right)
\nl
-12 (x-1)^2 x^2 (x+2) \text{Li}_2(x) \left(3 a^2 (x-2) x+2 a v \left(x^2-2 x-2\right)-v^2 \left(x^2-2 x+4\right)\right)
\nl
-384 v x^2 (a+v) \Re\text{Li}_2(x y)
\Bigg\}
\end{flalign}
\begin{flalign}
  \delta_{VG,r}^{T}(x,v,a)&=
\frac{1}{36 \pi  x \left(x^2-1\right)^3}
\Bigg\{
+24 v x \left(x^2-1\right)^3 \log \left(\frac{m_t^2}{\mu^2}\right)
+24 x \left(x^2-3\right) (a+v) \Re\text{Li}_2(x y)
\nl
+x \left(a \left(-3 \left(11 x^2-7\right) \left(x^2-1\right)^2-6 \pi  x \sqrt{4-x^2} \left(x^4-4 x^2+3\right)+2 \pi ^2 \left(3 x^5-6 x^4+4 x^2-3 x+1\right)\right)
\nll
+v \left(-3 \left(19 x^2-15\right) \left(x^2-1\right)^2-6 \pi  x \sqrt{4-x^2} \left(x^2-1\right)^2+2 \pi ^2 \left(3 x^5-6 x^4+4 x^2-3 x+1\right)\right)\right)
\nl
+24 \left(x^2-1\right) \sin ^{-1}\left(\frac{x}{2}\right) \left(a \left(\sqrt{4-x^2} \left(x^4-3 x^2-2\right)-\pi  x\right)+v x \left(x \sqrt{4-x^2} \left(3 x^2-4\right)-\pi \right)\right)
\nl
-24 x \left(x^2-6\right) (a+v) \sin ^{-1}\left(\frac{x}{2}\right)^2
-24 x^3 \left(x^2-1\right) (a+v) \log ^2(x)
\nl
-12 x \left(x^2-1\right) \log (x) \left(a \left(2 x^4-9 x^2-2\right)+v \left(2 x^4-3 x^2-2\right)\right)
\nl
-12 (x-1)^3 \left(x^3+3 x^2+x-1\right) (a+v) \log (1-x) \log (x)
\nl
+12 (x+1)^3 \left(x^3-3 x^2+x+1\right) (a+v) \log (x) \log (x+1)
\nl
+12 (x+1)^3 \left(x^3-3 x^2+x+1\right) (a+v) \text{Li}_2(-x)
-12 (x-1)^3 \left(x^3+3 x^2+x-1\right) (a+v) \text{Li}_2(x)
\Bigg\}
\end{flalign}
\begin{flalign}
  \delta_{VG,i}^{T}(x,v,a)&=-\frac{1}{3}(v+a)
\end{flalign}
\begin{flalign}
  \delta_{TG,r}^{T}(x,v,a)&=
\frac{1}{18 \pi  x \left(x^2-4\right) \left(x^2-1\right)^3}
\Bigg\{
+12 v x \left(x^2-4\right) \left(x^2-1\right)^3 \log \left(\frac{m_t^2}{\mu^2}\right)
\nl
+x \left(a \left(x^2-4\right) \left(2 x^8+\left(2 \pi ^2-41\right) x^6+\left(90-4 \pi ^2\right) x^4+\left(\pi ^2-65\right) x^2-6 \pi  \sqrt{4-x^2} x
\nlll
-3 \pi  \left(\sqrt{4-x^2}+4 \pi \right) x^5+3 \pi  \left(3 \sqrt{4-x^2}+4 \pi \right) x^3+14\right)
\nll
+v \left(\left(x^6-24 x^4+138 x^2-232\right) \left(x^2-1\right)^2-3 \pi  x \sqrt{4-x^2} \left(x^6-6 x^4+7 x^2-2\right)
\nlll
+\pi ^2 \left(4 x^8-12 x^7-8 x^6+60 x^5-41 x^4-48 x^3+32 x^2+16\right)\right)\right)
\nl
+6 \left(x^2-1\right) \sin ^{-1}\left(\frac{x}{2}\right) \left(a \left(x^2-4\right) \left(\sqrt{4-x^2} \left(2 x^4-3 x^2-4\right)-4 \pi  x\right)
\nll
+2 v x \left(x \sqrt{4-x^2} \left(3 x^4-16 x^2+13\right)+2 \pi \right)\right)
\nl
+12 x \sin ^{-1}\left(\frac{x}{2}\right)^2 \left(a \left(3 x^4-10 x^2-8\right)+v \left(-5 x^4+22 x^2-32\right)\right)
\nl
+6 x \left(x^4-5 x^2+4\right) \log (x) \left(a \left(x^4+8 x^2+2\right)-v x^2 \left(5 x^2+9\right)\right)
\nl
-6 (x-1)^3 \left(x^2+3 x+2\right) \log (1-x) \log (x) \left(a \left(4 x^4-9 x^3+4 x^2-5 x+2\right)+v x \left(2 x^3-x^2-8 x+3\right)\right)
\nl
-6 (x+1)^3 \left(x^2-3 x+2\right) \log (x) \log (x+1) \left(a \left(4 x^4+9 x^3+4 x^2+5 x+2\right)+v x \left(2 x^3+x^2-8 x-3\right)\right)
\nl
-6 (x+1)^3 \left(x^2-3 x+2\right) \text{Li}_2(-x) \left(a \left(4 x^4+9 x^3+4 x^2+5 x+2\right)+v x \left(2 x^3+x^2-8 x-3\right)\right)
\nl
-6 (x-1)^3 \left(x^2+3 x+2\right) \text{Li}_2(x) \left(a \left(4 x^4-9 x^3+4 x^2-5 x+2\right)+v x \left(2 x^3-x^2-8 x+3\right)\right)
\nl
-24 x \left(a \left(x^2-4\right)-v \left(x^4-5 x^2+7\right)\right) \Re\text{Li}_2(x y)
\Bigg\}
\end{flalign}
\begin{flalign}
  \delta_{TG,i}^{T}(x,v,a)&=-\frac{1}{3}(v+a)
\end{flalign}
\begin{flalign}
  \delta_{G2}^{T}(x,v,a)&=
\frac{1}{72 \pi  \left(x^2-4\right) \left(x^2-1\right)^2}
\Bigg\{
+a^2 \left(x^2-4\right) \left(4 x^6-15 x^4+126 x^2+3 \pi  \left(20-3 x^2\right) \sqrt{4-x^2} x
\nll
+\pi ^2 \left(15 x^3-6 x^2+15 x-2\right)-115\right)
\nl
+2 a v \left(57 x^6-330 x^4+453 x^2+3 \pi  \sqrt{4-x^2} \left(x^2-2\right)^2 x+\pi ^2 \left(5 x^4-2 x^3-25 x^2+6 x+10\right) x-180\right)
\nl
-v^2 \left(-4 x^8+31 x^6+6 x^4-341 x^2+3 \pi  \sqrt{4-x^2} \left(3 x^2-16\right) x^3+\pi ^2 \left(5 x^5-2 x^4+5 x^3+10 x^2-80 x+8\right)+308\right)
\nl
+6 \sin ^{-1}\left(\frac{x}{2}\right) \left(a^2 \left(x^2-4\right) \left(3 x \sqrt{4-x^2} \left(3 x^2-20\right)+4 \pi \right)+2 a v \left(x^2-2\right) \left(4 \pi -3 x \sqrt{4-x^2} \left(x^2-2\right)\right)
\nll
+v^2 x^2 \left(3 x \sqrt{4-x^2} \left(3 x^2-16\right)+4 \pi \right)\right)
-24 \sin ^{-1}\left(\frac{x}{2}\right)^2 \left(3 a^2 \left(x^2-4\right)+2 a v \left(3 x^2-14\right)+v^2 \left(3 x^2-16\right)\right)
\nl
-6 \left(x^2-4\right) \log (x) \left(a^2 \left(9 x^4-26 x^2+2\right)+10 a v \left(x^4+2 x^2-2\right)+v^2 \left(9 x^4-10 x^2-14\right)\right)
\nl
-24 (x-1)^2 (x+2) \log (1-x) \log (x) \left(3 a^2 (x-2) x+2 a v \left(x^2-2 x-2\right)-v^2 \left(x^2-2 x+4\right)\right)
\nl
+6 (x+2) \log ^2(x) \left(a^2 \left(3 x^4-24 x^3+39 x^2-2 x-8\right)+2 a v \left(x^4-8 x^3+11 x^2+6 x-10\right)
\nll
-v^2 \left(x^4-8 x^3+17 x^2-14 x+12\right)\right)
\nl
+24 (x-2) (x+1)^2 \log (x) \log (x+1) \left(3 a^2 x (x+2)+2 a v \left(x^2+2 x-2\right)-v^2 \left(x^2+2 x+4\right)\right)
\nl
+24 (x-2) (x+1)^2 \text{Li}_2(-x) \left(3 a^2 x (x+2)+2 a v \left(x^2+2 x-2\right)-v^2 \left(x^2+2 x+4\right)\right)
\nl
-12 (x-1)^2 (x+2) \text{Li}_2(x) \left(3 a^2 (x-2) x+2 a v \left(x^2-2 x-2\right)-v^2 \left(x^2-2 x+4\right)\right)
\nl
-384 v (a+v) \Re\text{Li}_2(x y)
\Bigg\}
\end{flalign}
\begin{flalign}
  \delta_{VG,r}^{F}(x,v,a)&=
\frac{1}{36 \pi  x \sqrt{4-x^2} \left(x^2-1\right)^3}
\Bigg\{
+24 v x \sqrt{4-x^2} \left(x^2-1\right)^3 \log \left(\frac{m_t^2}{\mu^2}\right)
\nl
+x \sqrt{4-x^2} \left(a \left(2 \pi ^2 \left(2 x^4-6 x^2+3\right)-3 (x-1)^2 \left(11 x^4+18 x^3-18 x^2-22 x+3\right)\right)
\nll
+v \left(2 \pi ^2 \left(6 x^4-2 x^2-5\right)-3 (x-1)^2 \left(19 x^4+34 x^3+22 x^2-38 x-45\right)\right)\right)
\nl
-120 x \sqrt{4-x^2} \left(x^2-2\right) (a+v) \sin ^{-1}\left(\frac{x}{2}\right)^2
+48 \left(x^2-1\right) \Im\text{Li}_2(y) \left(a \left(x^2-4\right)+v x^2\right)
\nl
+48 \left(x^2-1\right) \left(a \left(x^2-4\right)+v x^2\right) \Im\text{Li}_2((x-1) y)
-48 \left(x^2-1\right) \left(a \left(x^2-4\right)+v x^2\right) \Im\text{Li}_2(x y)
\nl
-24 \left(x^2-1\right) \left(a \left(x^2-4\right)+v x^2\right) \Im\text{Li}_2\left(
\frac{(x^2-1)y}{x}\right)
\nl
+12 x \sqrt{4-x^2} \left(x^2-1\right) \log (2-x) \left(a \left(x^4-5 x^2+4\right)+v \left(x^2-3\right) x^2\right)
\nl
-12 \left(x^2-1\right) \log (x) \left(a \left(x \sqrt{4-x^2} \left(x^4+3 x^2-4\right)-\pi  \left(x^2-4\right)\right)+v \left(x \sqrt{4-x^2} \left(x^2+1\right)-\pi \right) x^2\right)
\nl
+24 x^3 \sqrt{4-x^2} \left(x^2-1\right) (a+v) \log ^2(x)
+48 x \sqrt{4-x^2} \left(x^2-2\right) (a+v) \Re\text{Li}_2(x y)
\nl
+12 \left(x^2-1\right) \log (x+1) \left(a \left(x \sqrt{4-x^2} \left(x^2-1\right)^2-\pi  \left(x^2-4\right)\right)+v x \left(\sqrt{4-x^2} \left(x^4+6 x^2-7\right)-\pi  x\right)\right)
\nl
+\sin ^{-1}\left(\frac{x}{2}\right) \left(24 \left(x^2-1\right) \log (x) \left(a \left(x^2-4\right)+v x^2\right)+72 \left(x^2-1\right) \log (x+1) \left(a \left(x^2-4\right)+v x^2\right)
\nll
+12 \left(x^4-5 x^2+4\right) \left(a \left(x^4-3 x^2+4\right)+v \left(5-3 x^2\right) x^2\right)\right)
\nl
+24 x \sqrt{4-x^2} \left(x^2-1\right) \text{Li}_2(-x) \left(2 a x^2-5 a+6 v x^2+3 v\right)
\Bigg\}
\end{flalign}
\begin{flalign}
  \delta_{VG,i}^{F}(x,v,a)&=-\frac{1}{3}(v+a)
\end{flalign}
\begin{flalign}
  \delta_{TG,r}^{F}(x,v,a)&=
-\frac{1}{18 \pi  x \left(x^2-4\right) \left(x^2-1\right)^3}
\Bigg\{
-12 v x \left(x^2-4\right) \left(x^2-1\right)^3 \log \left(\frac{m_t^2}{\mu^2}\right)
\nl
-x \left(x^2-4\right) \left(a \left(x^8-2 \left(5+\pi ^2\right) x^6-20 x^5+63 x^4+8 x^3+\left(\pi ^2-68\right) x^2+12 x+14\right)
\nll
+v \left(\pi ^2 \left(4 x^4-x^2-4\right)+\left(2 x^6+4 x^5-29 x^4-66 x^3-25 x^2+68 x+58\right) (x-1)^2\right)\right)
\nl
+60 x \left(x^4-6 x^2+8\right) (a+v) \sin ^{-1}\left(\frac{x}{2}\right)^2
+48 \sqrt{4-x^2} \left(x^2-1\right) \Im\text{Li}_2(y) \left(v x^2-a\right)
\nl
+48 \sqrt{4-x^2} \left(x^2-1\right) \left(v x^2-a\right) \Im\text{Li}_2((x-1) y)
-48 \sqrt{4-x^2} \left(x^2-1\right) \left(v x^2-a\right) \Im\text{Li}_2(x y)
\nl
-24 \sqrt{4-x^2} \left(x^2-1\right) \left(v x^2-a\right) \Im\text{Li}_2\left(
\frac{(x^2-1)y}{x}\right)
+12 x^3 \left(x^2-4\right) \left(x^2-1\right)^2 (a+v) \log ^2(x)
\nl
-6 x \left(x^4-5 x^2+4\right) \log (2-x) \left(a \left(x^4-5 x^2+2\right)+v \left(x^2-2\right) x^2\right)
\nl
-6 \left(x^2-1\right) \log (x) \left(a \left(x^7-9 x^5+22 x^3+2 \pi  \sqrt{4-x^2}-8 x\right)+v \left(x^5-6 x^3-2 \pi  \sqrt{4-x^2}+8 x\right) x^2\right)
\nl
-12 \left(x^2-1\right) \log (x+1) \left(v x \left(2 x^4-10 x^2+\pi  \sqrt{4-x^2} x+8\right)-a \left(2 x^7-10 x^5+8 x^3+\pi  \sqrt{4-x^2}\right)\right)
\nl
+\sin ^{-1}\left(\frac{x}{2}\right) \left(24 \sqrt{4-x^2} \left(x^2-1\right) \log (x) \left(v x^2-a\right)+72 \sqrt{4-x^2} \left(x^2-1\right) \log (x+1) \left(v x^2-a\right)
\nll
+6 \sqrt{4-x^2} \left(x^4-5 x^2+4\right) \left(a \left(x^4-3 x^2+4\right)+v \left(5-3 x^2\right) x^2\right)\right)
\nl
+24 x \left(x^4-5 x^2+4\right) \text{Li}_2(-x) \left(a \left(x^4+x^2+1\right)-v \left(2 x^2+1\right)\right)
\nl
-24 x \left(x^4-6 x^2+8\right) (a+v) \Re\text{Li}_2(x y)
\Bigg\}
\end{flalign}
\begin{flalign}
  \delta_{TG,i}^{F}(x,v,a)&=-\frac{1}{3}(v+a)
\end{flalign}
\begin{flalign}
  \delta_{G2}^{F}(x,v,a)&=
\frac{1}{72 \pi  \sqrt{4-x^2} \left(x^2-1\right)^2}
\Bigg\{
-96 x (a+v)^2 \Im\text{Li}_2(y)
+96 x (a+v)^2 \Im\text{Li}_2(x y)
\nl
+(x-1) \sqrt{4-x^2} \left(3 a^2 \left(21 x^3-15 x^2+4 \left(8 x^2-9 x-21\right) \log (2)-31 x+25\right)
\nll
+2 a v \left(4 x^5+4 x^4-17 x^3-13 x^2-49 x+12 (3 x+11) \log (2)+71\right)
\nll
+3 v^2 \left(21 x^3-15 x^2+4 \left(8 x^2-x+3\right) \log (2)-31 x+25\right)\right)
\nl
-6 x^2 \sqrt{4-x^2} \log (2-x) \left(a^2 \left(x^2-2\right)-6 a v \left(x^2-6\right)+v^2 \left(x^2-2\right)\right)
-192 x (a+v)^2 \log (x) \sin ^{-1}\left(\frac{x}{2}\right)
\nl
-6 \sqrt{4-x^2} \log (x) \left(a^2 \left(7 x^4-16 x^3+x^2 (20-8 \log (2))+24 x-36\right)+2 a v \left(3 x^4-8 x^2 \log (2)-16 x+28\right)
\nll
+v^2 \left(7 x^4-16 x^3+x^2 (4-8 \log (2))-8 x+12\right)\right)
+72 x^2 \sqrt{4-x^2} (a+v)^2 \log ^2(x)
\Bigg\}
\ .
\end{flalign}

The $\delta^{\gamma}_{T,TG,G}(\xhat,\yhat)$ functions that appeared in
Eq.~(\ref{eq:fulltua2}) are:
\begin{flalign}
  \delta_{T}^{\gamma}(\xhat,\yhat)&=
\frac{1}{3 \pi }
\Bigg\{
+2 \log \left(\frac{m_t^2}{\mu ^2}\right)
-6
+\yhat
-\frac{4 \pi ^2}{3}
+\frac{16 (\yhat-1)}{\xhat^2}
+\frac{4 (\yhat-2) (\yhat-1)}{\xhat}
-\frac{2 (\yhat-1) \yhat (2 \yhat-1)}{\xhat \yhat-2}
\nl
+12 \sqrt{\frac{2}{\xhat}-1} \tan ^{-1}\left(\frac{1-\yhat}{\sqrt{\frac{2}{\xhat}-1}}\right)
+\left(\yhat^2+2 \yhat-10\right) \log (1-\yhat)
-2 \log ^2(1-\yhat)
\nl
-\frac{2}{\xhat^3}\left(\xhat^2-12 \xhat+16\right) \log \left(\frac{\xhat-2}{\xhat \yhat-2}\right)
-6 \log \left(\frac{2-\xhat}{\xhat (\yhat-2) \yhat+2}\right)
-\frac{\xhat \yhat (\yhat+2)+2}{\xhat} \log \left(\frac{\xhat (\yhat-2) \yhat+2}{2-\xhat \yhat}\right)
\nl
+4 \text{Li}_2\left(-\frac{\xhat (\yhat-1)}{\xhat-2}\right)
-2 \text{Li}_2\left(\frac{\xhat (\yhat-1)^2}{\xhat-2}\right)
\Bigg\}
\end{flalign}

\begin{flalign}
  \delta_{TG,r}^{\gamma}(\xhat,\yhat)&=
\frac{2}{9 \pi }
\Bigg\{
+4 \log \left(\frac{m_t^2}{\mu ^2}\right)
-11
+\frac{5 \pi ^2}{3}
-\frac{(2-\xhat) (1-\yhat) \left(\xhat^2 \yhat-2 \xhat \yhat-2 \xhat+8\right)}{\xhat^2 (2-\xhat \yhat)}
\nl
+4 \tan ^{-1}\left(\sqrt{\frac{2}{\xhat}-1}\right)^2
-8 \left(\sqrt{\frac{2}{\xhat}-1}-\tan ^{-1}\left(\sqrt{\frac{2}{\xhat}-1}\right)\right) \tan ^{-1}\left(\frac{1-\yhat}{\sqrt{\frac{2}{\xhat}-1}}\right)
\nl
-\frac{4}{\xhat^3} \left(\xhat^2-4 \xhat+4\right) \log \left(\frac{2-\xhat}{2-\xhat \yhat}\right)
-2 \log \left(\frac{\xhat \yhat^2}{2}\right) \log \left(\frac{1}{2} (2-\xhat (2-\yhat) \yhat)\right)
-4 (1-\yhat) \log \left(\frac{2-\xhat (2-\yhat) \yhat}{(1-\yhat) (2-\xhat \yhat)}\right)
\nl
+2 \left(-2 \text{Li}_2\left(\frac{\xhat \yhat}{2}\right)+\text{Li}_2\left(\frac{\xhat}{2}\right)-2 \text{Li}_2(\yhat)\right)
-8 \text{ReLi}_2\left(1-\frac{1}{2} \left(\xhat+i \sqrt{(2-\xhat) \xhat}\right) \yhat\right)
\Bigg\}
\end{flalign}

\begin{flalign}
  \delta_{TG,i}^{\gamma}(\xhat,\yhat)&=-\frac{4}{9}
\end{flalign}

\begin{flalign}
  \delta_{G}^{\gamma}(\xhat,\yhat)&=
\frac{4}{27 \pi }
\Bigg\{
+\frac{2 \pi ^2}{3}
-\frac{(2-\xhat) (1-\yhat) \left(3 \xhat^2 \yhat-4 \xhat \yhat-8 \xhat+16\right)}{\xhat^2 (2-\xhat \yhat)}
+\left(-2 \xhat+4 \log \left(\frac{\xhat}{2}\right)+4\right) \log (\yhat)
\nl
+\frac{2}{\xhat^3} (2-\xhat) \left(\xhat^3-\xhat^2+6 \xhat-8\right) \log \left(\frac{2-\xhat}{2-\xhat \yhat}\right)
+(1-\yhat) (3-\yhat) \log \left(\frac{\xhat (1-\yhat)}{2-\xhat \yhat}\right)
\nl
+4 \left(\text{Li}_2\left(\frac{\xhat \yhat}{2}\right)-\text{Li}_2\left(\frac{\xhat}{2}\right)-\text{Li}_2(\yhat)\right)
\Bigg\}
\ .
\end{flalign}

The $\delta_{S,SG,G3}(x)$ functions that appeared in Eq.~(\ref{eq:fulltuh2})
are [$\delta_S(x)$ also appeared in Eq.~(\ref{eq:fullST2})]:

\begin{flalign}
  \delta_S(x)&=
\frac{1}{9 \pi  x^2 \left(1-x^2\right)}
\Bigg\{
+36 x^2 \left(x^2-1\right) \log \left(\frac{m_t}{\mu }\right)
+\left(4 \pi ^2-51\right) x^2 \left(x^2-1\right)
+6 \left(5 x^4-7 x^2+2\right) \log (1-x)
\nl
-24 x^4 \log (x)
+24 x^2 \left(x^2-1\right) \log (1-x) \log (x)
+6 \left(5 x^4-7 x^2+2\right) \log (x+1)
\nl
+24 x^2 \left(x^2-1\right) \log (x) \log (x+1)
+48 x^2 \left(x^2-1\right) \text{Li}_2(-x)
+48 x^2 \left(x^2-1\right) \text{Li}_2(x)
\Bigg\}
\end{flalign}

\begin{flalign}
  \delta_{SG}(x)&=
\frac{1}{9 \pi  \left(x^2-1\right)^2}
\Bigg\{
-36 \left(x^2-1\right)^2 \log \left(\frac{m_t}{\mu }\right)
-6 \left(\frac{\sqrt{4-x^2}}{x} \left(x^4-6 x^2+8\right)+2 \pi \right) \sin ^{-1}\left(\frac{x}{2}\right)
\nl
-36 \sin ^{-1}\left(\frac{x}{2}\right)^2
-6 x^2 \left(5 x^2+2\right) \log (x)
-24 \log (1-x) \log (x)
+12 \log ^2(x)
-24 \log (x) \log (x+1)
\nl
-24 \text{Li}_2(-x)
-24 \text{Li}_2(x)
+24 \Re\text{Li}_2\left(\frac{(x^2-1)y}{x}\right)-3 \pi  \sqrt{4-x^2} \left(x^2-2\right) x+3 \left(x^4+8 x^2-9\right) x^2+5 \pi ^2
\Bigg\}
\end{flalign}

\begin{flalign}
  \delta_{G3}(x)&=
-\frac{1}{36 \pi \left(1-x^2\right)^2}
\Bigg\{ +271 -620 x^2 +342 x^4 +8 x^6 -x^8 -156 \pi  x \sqrt{4-x^2}
+30 \pi  x^3 \sqrt{4-x^2}
\nl
+36 x \left(26-5 x^2\right) \sqrt{4-x^2} \sin ^{-1}\left(\frac{x}{2}\right)
-12 \left(9 x^4+76 x^2-8\right) \log (x)
\Bigg\}
\ .
\end{flalign}

The $\delta_{ST}(x)$ and $\delta_{TT}^{(i)}$ functions that appeared in 
Eq.~(\ref{eq:fullST2}) are:
  \begin{flalign}
    \delta_{ST}(x)=&
   -\frac{1}{9 \pi }\left[
     6 \log \left(\frac{m_t^2}{\mu ^2}\right)+\frac{24
   \left(2-x^2\right) x^2 \text{Li}_2(-x)}{\left(1-x^2\right)^2}+24
   \text{Li}_2(x)-\frac{3 \left(7 x^2+6 x+5\right)}{(x+1)^2}+\frac{2 \pi
   ^2}{\left(1-x^2\right)^2}
   \right.\nonumber\\&\left.-\frac{6 \left(3-x^2\right) \log
   (x+1)}{1-x^2}+18 \log (1-x)
   \right]
  \end{flalign}
  \begin{flalign}
    \delta_{TT}^{(1)}(x)=&
    -\frac{1}{9 \pi  x^2 \left(x^2-1\right)^2}
    \left[-6 x^2 \left(x^2-1\right)^2 \log \left(\frac{m_t^2}{\mu
    ^2}\right)+12 x \left(4 x^3+7 x^2+2 x+3\right) (x-1)^2 \text{Li}_2(x)
    \right.\nonumber\\&\left.
    +12 x (x+1)^2 \left(4 x^3-7 x^2+2 x-3\right) \text{Li}_2(-x)+6 \left(x^6-3
    x^2+2\right) \log (1-x)+6 \left(x^6-3 x^2+2\right) \log (x+1)+x^2
    \right.\nonumber\\&\left.
    \left(3 \left(x^4-22 x^2+21\right)+4 \pi ^2 \left(x^4-2 x^2-3\right)\right)-24 x^2
    \left(x^4-3 x^2-3\right) \log (x)
    \right.\nonumber\\&\left.
    +12 x \left(2 x^3+3 x^2+3\right) (x-1)^2
    \log (1-x) \log (x)
    +12 x (x+1)^2 \left(2 x^3-3 x^2-3\right) \log (x) \log
    (x+1)\right]
    \\
    \delta_{TT}^{(2)}(x)=&
    -\frac{1}{9 \pi \left(x^2-1\right)^2}\left[
    -6 \left(x^2-1\right)^2 \log \left(\frac{m_t^2}{\mu
    ^2}\right)-6 (x+1)^2 \left(x^3-10 x^2+13 x-8\right) \text{Li}_2(-x)
    \right.\nonumber\\&\left.
    +6 (x-1)^2 \left(x^3+10 x^2+13 x+8\right) \text{Li}_2(x)-5 x^6+30 x^4+4 \pi ^2
    \left(x^4+1\right)-9 x^2+18 \left(x^2-1\right)^2 \log (1-x)
    \right.\nonumber\\&\left.
    +18 \left(x^2-1\right)^2 \log (x+1)-12 \left(5 x^4+x^2\right) \log (x)+6 (x-1)^2
    \left(x^3+6 x^2+5 x+4\right) \log (1-x) \log (x)
    \right.\nonumber\\&\left.
    +6 \left(-x^5+4 x^4+6 x^3+3
    x+4\right) \log (x) \log (x+1)-16\right]
    \ .
  \end{flalign}
\end{widetext}

\section{Numerical results for flavor-changing decays}
\label{sec:apptable}

The numerical values of $\Gamma_{ij}$ defined in Eq.~(\ref{eq:numgamma}) are
given in Tables~\ref{tab:fcnc1}--\ref{tab:fcnc6}.  In these tables we show the
contributions from two-fermion operators, as well as their interferences with
the V-V four-fermion operators. To save space we do not include the operator
$O_{\varphi q}^{(3,1+3)}$, whose contribution is the same as $O_{\varphi
q}^{(1,1+3)}$ but with a different sign, and the operator $O_{lq}^{(3,1+3)}$,
whose contribution is the same as $O_{lq}^{(1,1+3)}$.  The percentage number
under each entry represents the NLO correction (a dash implies the presented
value vanishes at LO).  

We do not display $\Gamma_{ij}$ when both $i$ and $j$ correspond to V-V, S-S
and T-T operators.  These contributions are simply
\begin{flalign}
  \Gamma_{t\to ue^+e^-}=&
  2.9\times 10^{-6} \mathrm{GeV}
  \left(\left|C_{lq}^{(1,1+3)}+C_{lq}^{(3,1+3)}\right|^2
  \right.\nonumber\\&\left.
  +\left|C_{qe}^{(1+3)}\right|^2+
  \left|C_{lu}^{(1+3)}\right|^2+\left|C_{eu}^{(1+3)}\right|^2\right)
  \nonumber\\
  &+8.2\times 10^{-7} \mathrm{GeV}
  \left(\left|C_{lequ}^{(1,13)}\right|^2
  +\left|C_{lequ}^{(1,31)}\right|^2\right)
  \nonumber\\
  &+3.5\times 10^{-5} \mathrm{GeV} \left(\left|C_{lequ}^{(3,13)}\right|^2
  +\left|C_{lequ}^{(3,31)}\right|^2\right)
  \ .
\end{flalign}
The NLO corrections in these three terms are about $-8\%$, $1\%$ and $-8\%$
respectively.  The interference between S-S and T-T operators is proportional
to $\cos\theta$ and vanishes upon phase space integration.  There is no other
interference between any two four-fermion operators.

\begin{table*}[h!]
  \centering
\[
\begin{array}{c|cccccccc}
  \Gamma_{ij}(\mathrm{GeV}) & \Re\left(C_{\varphi q}^{(1,1+3)}\right) &
  \Im\left(C_{\varphi q}^{(1,1+3)}\right) & \Re\left(C_{uW}^{(13)}\right) &
  \Im\left(C_{uW}^{(13)}\right) & \Re\left(C_{uB}^{(13)}\right) &
  \Im\left(C_{uB}^{(13)}\right) & \Re\left(C_{uG}^{(13)}\right) &
  \Im\left(C_{uG}^{(13)}\right) \\\hline
 \Re\left(C_{\varphi q}^{(1,1+3)}\right) & \underset{-8\%}{1.9\times 10^{-5}} & \underset{\text{---}}{0.} & \underset{-8\%}{-6.2\times 10^{-5}} & \underset{-8\%}{1.6\times 10^{-7}} & \underset{-8\%}{1.8\times 10^{-5}} & \underset{-8\%}{1.6\times 10^{-7}} & \underset{\text{---}}{2.9\times 10^{-6}} & \underset{\text{---}}{-2.2\times 10^{-6}} \\
 \Im\left(C_{\varphi q}^{(1,1+3)}\right) & \underset{}{} & \underset{-8\%}{1.9\times 10^{-5}} & \underset{-8\%}{-1.6\times 10^{-7}} & \underset{-8\%}{-6.2\times 10^{-5}} & \underset{-8\%}{-1.6\times 10^{-7}} & \underset{-8\%}{1.8\times 10^{-5}} & \underset{\text{---}}{2.2\times 10^{-6}} & \underset{\text{---}}{2.9\times 10^{-6}} \\
 \Re\left(C_{uW}^{(13)}\right) & \underset{}{} & \underset{}{} & \underset{-9\%}{7.6\times 10^{-5}} & \underset{\text{---}}{0.} & \underset{-9\%}{-3.6\times 10^{-5}} & \underset{-9\%}{-4.9\times 10^{-7}} & \underset{\text{---}}{-6.1\times 10^{-6}} & \underset{\text{---}}{5.5\times 10^{-6}} \\
 \Im\left(C_{uW}^{(13)}\right) & \underset{}{} & \underset{}{} & \underset{}{} & \underset{-9\%}{7.6\times 10^{-5}} & \underset{-9\%}{4.9\times 10^{-7}} & \underset{-9\%}{-3.6\times 10^{-5}} & \underset{\text{---}}{-5.5\times 10^{-6}} & \underset{\text{---}}{-6.1\times 10^{-6}} \\
 \Re\left(C_{uB}^{(13)}\right) & \underset{}{} & \underset{}{} & \underset{}{} & \underset{}{} & \underset{-9\%}{9.1\times 10^{-6}} & \underset{\text{---}}{0.} & \underset{\text{---}}{1.4\times 10^{-6}} & \underset{\text{---}}{-1.2\times 10^{-6}} \\
 \Im\left(C_{uB}^{(13)}\right) & \underset{}{} & \underset{}{} & \underset{}{} & \underset{}{} & \underset{}{} & \underset{-9\%}{9.1\times 10^{-6}} & \underset{\text{---}}{1.2\times 10^{-6}} & \underset{\text{---}}{1.4\times 10^{-6}} \\
 \Re\left(C_{uG}^{(13)}\right) & \underset{}{} & \underset{}{} & \underset{}{} & \underset{}{} & \underset{}{} & \underset{}{} & \underset{\text{---}}{6.8\times 10^{-8}} & \underset{\text{---}}{0.} \\
 \Im\left(C_{uG}^{(13)}\right) & \underset{}{} & \underset{}{} & \underset{}{} & \underset{}{} & \underset{}{} & \underset{}{} & \underset{}{} & \underset{\text{---}}{6.8\times 10^{-8}} \\
\end{array}
\]
  \caption{
Contributions from two-fermion operators that involve a left-handed light quark.
  }
  \label{tab:fcnc1}
\end{table*}

\begin{table*}
  \centering
  \[
\begin{array}{c|cccc}
 \Gamma_{ij}(\mathrm{GeV}) & \Re\left(C_{lq}^{(1,1+3)}\right) &
 \Im\left(C_{lq}^{(1,1+3)}\right) & \Re\left(C_{qe}^{(1+3)}\right) &
 \Im\left(C_{qe}^{(1+3)}\right) \\\hline
 \Re\left(C_{\varphi q}^{(1,1+3)}\right) & \underset{-12\%}{-3.5\times 10^{-7}} & \underset{-8\%}{-2.3\times 10^{-6}} & \underset{-12\%}{2.8\times 10^{-7}} & \underset{-8\%}{1.8\times 10^{-6}} \\
 \Im\left(C_{\varphi q}^{(1,1+3)}\right) & \underset{-8\%}{2.3\times 10^{-6}} & \underset{-12\%}{-3.5\times 10^{-7}} & \underset{-8\%}{-1.8\times 10^{-6}} & \underset{-12\%}{2.8\times 10^{-7}} \\
 \Re\left(C_{uW}^{(13)}\right) & \underset{-7\%}{-3.3\times 10^{-6}} & \underset{-8\%}{3.8\times 10^{-6}} & \underset{-7\%}{-1.4\times 10^{-6}} & \underset{-8\%}{-3.\times 10^{-6}} \\
 \Im\left(C_{uW}^{(13)}\right) & \underset{-8\%}{-3.8\times 10^{-6}} & \underset{-7\%}{-3.3\times 10^{-6}} & \underset{-8\%}{3.\times 10^{-6}} & \underset{-7\%}{-1.4\times 10^{-6}} \\
 \Re\left(C_{uB}^{(13)}\right) & \underset{-7\%}{-1.9\times 10^{-6}} & \underset{-8\%}{-1.1\times 10^{-6}} & \underset{-7\%}{-2.5\times 10^{-6}} & \underset{-8\%}{8.7\times 10^{-7}} \\
 \Im\left(C_{uB}^{(13)}\right) & \underset{-8\%}{1.1\times 10^{-6}} & \underset{-7\%}{-1.9\times 10^{-6}} & \underset{-8\%}{-8.7\times 10^{-7}} & \underset{-7\%}{-2.5\times 10^{-6}} \\
 \Re\left(C_{uG}^{(13)}\right) & \underset{\text{---}}{2.6\times 10^{-7}} & \underset{\text{---}}{-2.5\times 10^{-8}} & \underset{\text{---}}{6.4\times 10^{-9}} & \underset{\text{---}}{2.3\times 10^{-7}} \\
 \Im\left(C_{uG}^{(13)}\right) & \underset{\text{---}}{2.5\times 10^{-8}} & \underset{\text{---}}{2.6\times 10^{-7}} & \underset{\text{---}}{-2.3\times 10^{-7}} & \underset{\text{---}}{6.4\times 10^{-9}} \\
\end{array}
\]
  \caption{
Interference between two-fermion operators and V-V four-fermion operators that involve a left-handed light quark.
  } \label{tab:fcnc3}
\end{table*}

\begin{table*}
  \centering
\[
\begin{array}{c|cccccccc}
 \Gamma_{ij}(\mathrm{GeV}) & \Re\left(C_{\varphi u}^{(1+3)}\right) &
 \Im\left(C_{\varphi u}^{(1+3)}\right) & \Re\left(C_{uW}^{(31)}\right) &
 \Im\left(C_{uW}^{(31)}\right) & \Re\left(C_{uB}^{(31)}\right) &
 \Im\left(C_{uB}^{(31)}\right) & \Re\left(C_{uG}^{(31)}\right) &
 \Im\left(C_{uG}^{(31)}\right) \\\hline
 \Re\left(C_{\varphi u}^{(1+3)}\right) & \underset{-8\%}{1.9\times 10^{-5}} & \underset{\text{---}}{0.} & \underset{-8\%}{-6.2\times 10^{-5}} & \underset{-8\%}{-1.6\times 10^{-7}} & \underset{-8\%}{1.8\times 10^{-5}} & \underset{-8\%}{-1.6\times 10^{-7}} & \underset{\text{---}}{-1.8\times 10^{-6}} & \underset{\text{---}}{-9.2\times 10^{-7}} \\
 \Im\left(C_{\varphi u}^{(1+3)}\right) & \underset{}{} & \underset{-8\%}{1.9\times 10^{-5}} & \underset{-8\%}{-1.6\times 10^{-7}} & \underset{-8\%}{6.2\times 10^{-5}} & \underset{-8\%}{-1.6\times 10^{-7}} & \underset{-8\%}{-1.8\times 10^{-5}} & \underset{\text{---}}{-9.2\times 10^{-7}} & \underset{\text{---}}{1.8\times 10^{-6}} \\
 \Re\left(C_{uW}^{(31)}\right) & \underset{}{} & \underset{}{} & \underset{-9\%}{7.6\times 10^{-5}} & \underset{\text{---}}{0.} & \underset{-9\%}{-3.6\times 10^{-5}} & \underset{-9\%}{4.9\times 10^{-7}} & \underset{\text{---}}{3.1\times 10^{-6}} & \underset{\text{---}}{1.9\times 10^{-6}} \\
 \Im\left(C_{uW}^{(31)}\right) & \underset{}{} & \underset{}{} & \underset{}{} & \underset{-9\%}{7.6\times 10^{-5}} & \underset{-9\%}{-4.9\times 10^{-7}} & \underset{-9\%}{-3.6\times 10^{-5}} & \underset{\text{---}}{-1.9\times 10^{-6}} & \underset{\text{---}}{3.1\times 10^{-6}} \\
 \Re\left(C_{uB}^{(31)}\right) & \underset{}{} & \underset{}{} & \underset{}{} & \underset{}{} & \underset{-9\%}{9.1\times 10^{-6}} & \underset{\text{---}}{0.} & \underset{\text{---}}{-1.3\times 10^{-6}} & \underset{\text{---}}{-9.6\times 10^{-7}} \\
 \Im\left(C_{uB}^{(31)}\right) & \underset{}{} & \underset{}{} & \underset{}{} & \underset{}{} & \underset{}{} & \underset{-9\%}{9.1\times 10^{-6}} & \underset{\text{---}}{9.6\times 10^{-7}} & \underset{\text{---}}{-1.3\times 10^{-6}} \\
 \Re\left(C_{uG}^{(31)}\right) & \underset{}{} & \underset{}{} & \underset{}{} & \underset{}{} & \underset{}{} & \underset{}{} & \underset{\text{---}}{5.3\times 10^{-8}} & \underset{\text{---}}{0.} \\
 \Im\left(C_{uG}^{(31)}\right) & \underset{}{} & \underset{}{} & \underset{}{} & \underset{}{} & \underset{}{} & \underset{}{} & \underset{}{} & \underset{\text{---}}{5.3\times 10^{-8}} \\
\end{array}
\]
  \caption{
Contributions from two-fermion operators that involve a right-handed light quark.
  } \label{tab:fcnc4}
\end{table*}

\begin{table*}
  \centering
  \[
\begin{array}{c|cccc}
 \Gamma_{ij}(\mathrm{GeV}) & \Re\left(C_{lu}^{(1+3)}\right) &
 \Im\left(C_{lu}^{(1+3)}\right) & \Re\left(C_{eu}^{(1+3)}\right) &
 \Im\left(C_{eu}^{(1+3)}\right) \\\hline
 \Re\left(C_{\varphi u}^{(1+3)}\right) & \underset{-12\%}{-3.5\times 10^{-7}} & \underset{-8\%}{-2.3\times 10^{-6}} & \underset{-12\%}{2.8\times 10^{-7}} & \underset{-8\%}{1.8\times 10^{-6}} \\
 \Im\left(C_{\varphi u}^{(1+3)}\right) & \underset{-8\%}{2.3\times 10^{-6}} & \underset{-12\%}{-3.5\times 10^{-7}} & \underset{-8\%}{-1.8\times 10^{-6}} & \underset{-12\%}{2.8\times 10^{-7}} \\
 \Re\left(C_{uW}^{(31)}\right) & \underset{-7\%}{-3.3\times 10^{-6}} & \underset{-8\%}{3.8\times 10^{-6}} & \underset{-7\%}{-1.4\times 10^{-6}} & \underset{-8\%}{-3.\times 10^{-6}} \\
 \Im\left(C_{uW}^{(31)}\right) & \underset{-8\%}{3.8\times 10^{-6}} & \underset{-7\%}{3.3\times 10^{-6}} & \underset{-8\%}{-3.\times 10^{-6}} & \underset{-7\%}{1.4\times 10^{-6}} \\
 \Re\left(C_{uB}^{(31)}\right) & \underset{-7\%}{-1.9\times 10^{-6}} & \underset{-8\%}{-1.1\times 10^{-6}} & \underset{-7\%}{-2.5\times 10^{-6}} & \underset{-8\%}{8.7\times 10^{-7}} \\
 \Im\left(C_{uB}^{(31)}\right) & \underset{-8\%}{-1.1\times 10^{-6}} & \underset{-7\%}{1.9\times 10^{-6}} & \underset{-8\%}{8.7\times 10^{-7}} & \underset{-7\%}{2.5\times 10^{-6}} \\
 \Re\left(C_{uG}^{(31)}\right) & \underset{\text{---}}{7.\times 10^{-8}} & \underset{\text{---}}{2.1\times 10^{-7}} & \underset{\text{---}}{1.6\times 10^{-7}} & \underset{\text{---}}{4.3\times 10^{-8}} \\
 \Im\left(C_{uG}^{(31)}\right) & \underset{\text{---}}{2.1\times 10^{-7}} & \underset{\text{---}}{-6.8\times 10^{-8}} & \underset{\text{---}}{4.3\times 10^{-8}} & \underset{\text{---}}{-1.6\times 10^{-7}} \\
\end{array}
\]
  \caption{
Interference between two-fermion operators and V-V four-fermion
operators that involve a right-handed light quark.
  } \label{tab:fcnc6}
\end{table*}

\bibliography{bib}
\bibliographystyle{unsrt}

\end{document}